\documentclass[pdflatex,sn-mathphys-num]{sn-jnl}

\usepackage{graphicx}%
\usepackage{multirow}%
\usepackage{amsmath,amssymb,amsfonts}%
\usepackage{amsthm}%
\usepackage{mathrsfs}%
\usepackage{mathalpha}
\usepackage[title]{appendix}%
\usepackage{xcolor}%
\usepackage{textcomp}%
\usepackage{manyfoot}%
\usepackage{booktabs}%
\usepackage{algorithm}%
\usepackage{algorithmicx}%
\usepackage{algpseudocode}%
\usepackage{listings}%
\usepackage{longtable}%
\usepackage{array}%
\usepackage{multirow}%
\usepackage{gensymb}%
\usepackage[inkscapelatex=false]{svg}%

\theoremstyle{thmstyleone}%

\theoremstyle{thmstyletwo}%

\theoremstyle{thmstylethree}%

\begin{document}

\title[Article Title]{Urban Heat MiniCubes: An AI-Ready dataset for urban heat research}

\author*[1]{\fnm{Jonathan} \sur{Starfeldt}}\email{jonstar@umd.edu}
\author[1]{\fnm{Maria J.} \sur{Molina}}
\author[2]{\fnm{Alexander} \sur{Kerr}}
\author[2]{\fnm{Adam} \sur{Yang}}
\author[3]{\fnm{Thomas R. H.} \sur{Holmes}}
\author[4]{\fnm{Christopher R.} \sur{Hain}}

\affil[1]{Department of Atmospheric and Oceanic Science, University of Maryland, College Park, MD, USA}
\affil[2]{Department of Computer Science, University of Maryland, College Park, MD, USA}
\affil[3]{NASA Goddard Space Flight Center, Greenbelt, MD, USA}
\affil[4]{NASA Marshall Space Flight Center, Huntsville, AL, USA}

\abstract{Urban heat is amplified by impermeable surfaces and heterogeneous built environments, yet street-level variability remains difficult to quantify because multi-sensor observations are rarely available in consistent, analysis-ready form at the necessary spatiotemporal scales. We present ``Urban Heat MiniCubes,'' a publicly available, FAIR-oriented dataset designed for machine learning applications in urban heat research. The dataset provides harmonized $90\times90$ km gridded data cubes for 48 cities in the Western Hemisphere spanning 2022-2023, with variables reprojected and collocated to a common grid to reduce preprocessing (e.g., reprojection, resampling, and spatiotemporal alignment). Urban Heat MiniCubes includes two complementary modalities: (i) higher-spatial-resolution, lower-frequency observations from Landsat 8/9 (e.g., surface reflectances) and Sentinel-1 (e.g., synthetic aperture radar backscatter), and (ii) higher-temporal-frequency, coarser observations from GOES-R (e.g., longwave infrared brightness temperatures) and a microwave land surface temperature product. We document variables and metadata and provide technical assessment using inter-variable analyses and autoencoder-based reconstruction-error summaries across pixel classes (e.g., water and cloud). Potential use cases and limitations are also discussed.
}


\maketitle

\section{Background \& Summary} \label{Intro}

Urban heat refers to the amplification of heat in built environments, influenced by factors such as vegetation cover and urban density \cite{kumar2024urban, qi2022HeatDecision, memon2009investigation}. With over half of the world's population and over eighty percent of the United States' population living in cities, monitoring urban heat is critical for public health and well-being \cite{USCensus, rizwan2008review}. Disadvantaged and vulnerable populations are often concentrated in areas where urban heat is amplified, making urban heat an environmental justice issue \cite{wolch2014urban, chakraborty2023residential, mitchell2015landscapes}. Although urban heat has been extensively studied, near-real-time monitoring at the neighborhood or street level is not widely available to the public. Artificial intelligence (AI) presents an opportunity to advance both the science and monitoring of urban heat \cite{molina2023review, mohamed2024AIHeat, ahmed2025AIHeat}, but harmonizing remote sensing data across sources is complex, requiring considerable domain expertise and data preparation \cite{young2017survival, wang2025AIReady, scheffler2020harmonization}. ``Urban Heat MiniCubes'' seeks to fill this need for the scientific community.

Urban Heat MiniCubes adheres to the FAIR data principles (Findable, Accessible, Interoperable, and Reusable), a widely used open-science framework for sharing data to enable its reuse \cite{wilkinson2016fair, verhulst2025AIReady}. These principles aim to reduce the time spent on data discovery and preprocessing, and increase the time spent on scientific analysis. Urban Heat MiniCubes is also ``AI-ready,'' which emphasizes the following four data categories: preparation, quality, documentation, and access \cite{esipaiready, wang2025AIReady}. Together, these standards can make it easier for scientists to integrate datasets into AI workflows. Few urban heat datasets currently meet both FAIR and AI-ready criteria. Newly developed foundation models can provide transferable latent representations that support reuse across tasks, including in scientific settings where they map heterogeneous physical variables into a shared representation; however, these representations are often not directly human-interpretable \cite{brown2025foundations, feng2025tessera}.

While dense networks of near-surface weather conditions (e.g., urban meteorological networks, i.e., micronets) can provide detailed urban temperature information, they can be costly to deploy and maintain, and can face practical siting and operational constraints \cite{mcpherson2007statewide, brotzge2020technical}. As an alternative, remotely sensed land surface temperature (LST) is often correlated with near-surface air temperature and offers a path forward to estimate urban heat experienced by humans at high-resolution \cite{good2017spatiotemporal, do2022comparison, naserikia2023land}. Thermal infrared (TIR) satellite instruments measure electromagnetic radiation emitted from Earth’s surface at specific wavelengths, from which brightness temperature is derived by the Planck function. LST retrieval then accounts for surface emissivity (often estimated using normalized difference vegetation index, i.e., NDVI, see subsection \ref{Landsat Var Selection}) and related parameters, with additional terrain-related corrections sometimes applied in complex topography \cite{LandsatCalVal, wang2019comparison, malakar2018LST, parastatidis2017LST}. However, differences in measured wavelength bands and spatiotemporal resolutions across satellite instruments pose additional challenges for data harmonization.

\begin{figure}[ht]
    \centering
    \includegraphics[width=1.0\linewidth]{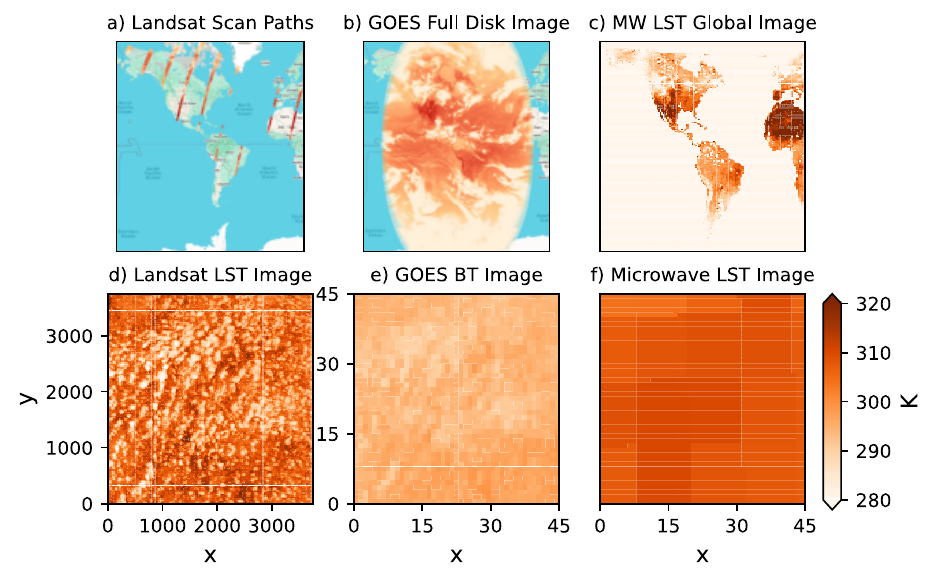}
    \caption{Illustration of different satellite sensor measurements. a) LST imagery from Landsat 8, a polar orbiting satellite, over the approximate area of the GOES East satellite. Note the limited spatial coverage of Landsat 8 intersecting land (water-only swaths omitted) during a single day. b) A single ABI band 14 image from GOES-16, a geostationary satellite. c) A global microwave LST image, cropped to the area of figures a and b. d) Landsat 8 LST image. Clouds appear colder (white) because the infrared radiation used to measure Landsat LST is attenuated by clouds. e) GOES Band 14 BT image. Temperature values appear colder than figures d and f because they are retrieved at the top of the atmosphere. f) Microwave LST image. The spatial resolution is much coarser than that of Landsat 8/9 LST, but microwave radiation is not attenuated by clouds. All imagery is taken between 16:10 and 16:15 UTC on 13 June 2022, except for (a), which is from the full day of 13 June 2022. Panels d-f are over Atlanta, Georgia, USA.}
    \label{satellite ref fig}
\end{figure}

High-spatial-temporal-resolution urban heat estimation is possible using multiple data modalities. One such modality includes polar-orbiting satellites, such as Landsat 8/9, which measure at high spatial resolution (e.g., 30m) but infrequently, with intervals of 12 hours \cite[e.g., VIIRS;][]{murphy2006visible} or more between local observations \cite{wulder2022fifty, benali2012estimating} (Fig. \ref{satellite ref fig}). Geostationary satellites measure at high temporal resolution (e.g., every 10 minutes) but at coarser spatial resolutions (e.g., 2km), which are insufficient for neighborhood-scale monitoring \cite{lindsey2024geoxo} (Fig. \ref{satellite ref fig}). Specific instruments aboard satellites measure infrared wavelengths that cannot penetrate clouds, limiting measurements to cloud-free days \cite{ni2024review}. Microwave sensors offer ``all-weather'' capabilities, but typically at coarser spatial resolutions \cite{prigent2016toward, jones2007satellite} (Fig. \ref{satellite ref fig}). Urban Heat MiniCubes unifies these diverse data modalities onto a common spatiotemporal grid for 48 cities across the Western Hemisphere. This AI-ready dataset aims to democratize access to remote sensing data and enable all-weather urban heat monitoring. The following sections describe its technical specifications, key properties, and potential applications.


\section{Data Overview}

\subsection{Landsat} 

\subsubsection{Landsat Overview} \label{Landsat overview}

The Landsat program is a long-running series of polar-orbiting passive remote-sensing systems that capture high-resolution images of naturally reflected or emitted radiation from the Earth's surface \cite{wulder2022fifty}. There have been nine Landsat satellites to date, the first of which was launched on 3 July 1972. We use data from Landsat 8 and 9 in our dataset, which were launched on 11 February 2013 and 27 September 2021, respectively. Both satellites are still in operation today. We use Landsat 8/9 due to the presence of two relevant instruments on board: the Operational Land Imager (OLI) and the Thermal Infrared Sensor (TIRS). 

OLI measures reflected solar radiation at the top of the atmosphere (TOA) at visible and shortwave infrared (SWIR) wavelengths at 30 m resolution \cite{LandsatHandbook}. TIRS has two TIR bands that measure brightness temperature at 100 m resolution \cite{LandsatHandbook}. Landsat 8/9 are in continuous sun-synchronous orbits, such that the satellites travel from north to south over the sunlit side of Earth and from south to north over the nighttime side, located approximately a half orbital period apart. Each orbit takes 99 minutes and passes the equator in its descending (i.e., north to south) direction at 10:00 am ± 15 minutes \cite{LandsatHandbook}. The full orbital cycle of each satellite repeats every 16 days, so the revisit time for Landsat imagery in Urban Heat MiniCubes is 8 days, due to the inclusion of both Landsat 8 and 9.

Landsat’s products are typically distributed in the Universal Transverse Mercator (UTM) system. In UTM coordinates, Earth is split into 60 longitudinal zones, numbered 1-60, each spanning six degrees of longitude, and further divided into northern and southern hemispheres to distinguish coordinates relative to the equator. The first zone is 180-174 degrees west longitude, with each successive zone being six degrees to the east. The UTM system uses meters as units; easting for the x-coordinate, and northing for the y-coordinate.

\subsubsection{Landsat Variable Selection} \label{Landsat Var Selection}

Observations from OLI and TIRS are used as inputs to multiple algorithms: the Landsat Surface Temperature (ST; commonly referred to as LST) algorithm, the Landsat Cloud Mask algorithm, and the Landsat Surface Reflectance algorithm. The outputs of these algorithms are atmospherically corrected Level 2 (L2) data products. Landsat LST is widely used in studies of urban heat, due to its high spatial resolution, which enables street-level monitoring of the thermal environment \cite{imran2021impact, zhao2020assessing, naserikia2023land}. Surface reflectances do not directly measure emitted thermal radiation; instead, they support the calculation of surface properties (e.g., vegetation, built-up, and albedo) that explain spatial variation in Landsat LST. As discussed in Section \ref{Intro}, infrared radiation, used to derive Landsat LST, is attenuated by clouds. Thus, pixels with clouds measure a ``cooler'' temperature than the actual surface temperature. We incorporate the Landsat Cloud Mask to monitor the impact of cloud contamination on Landsat LST.

The Landsat 8/9 (Collection 2) LST contained in our dataset uses a single-channel algorithm applied to TIRS Band-10 brightness temperatures \cite{LandsatCalVal, wang2019comparison}. Other auxiliary inputs include the Landsat Cloud-Mask product (CFMask), atmospheric profile variables (e.g., geopotential height, air temperature, and specific humidity) from the Goddard Earth Observing System Version 5 Forward Processing for Instrument Teams (GOES-5 FP-IT) reanalysis, and the Advanced Spaceborne Thermal Emission and Reflection Radiometer (ASTER) Global Emissivity Database (GED) and ASTER NDVI \cite{LandsatCalVal}. The inputs are used to run an atmospheric transmission model, which estimates surface temperature.

The Landsat 8/9 Cloud Mask is generated using the CFMask (C version of Fmask) algorithm, which applies decision-tree logic to OLI surface reflectances (including Band 9 from OLI, the cirrus band) to make normalized difference indices (Equation \ref{NDEquation}), used for land-cover classification, TIRS brightness temperatures to assess thermal contrast and cloud presence, and a digital elevation model to account for topographic effects \cite{LandsatCalVal, foga2017cloud}. The algorithm classifies each pixel as cloud, cloud shadow, water, snow, or clear. Cloud shadows are located by estimating cloud height and projecting the shadow position using the solar azimuth angle at the time of acquisition \cite{LandsatCalVal}. A confidence level of high, medium, or low is then assigned to each cloud pixel based on set temperature thresholds and surface characteristics. The outputs of both the LST and CFMask algorithms are delivered at 30 m resolution.

Landsat 8/9 L2 surface reflectances are generated using the Land Surface Reflectance Code (LaSRC) algorithm. LaSRC atmospherically corrects TOA reflectance measurements to achieve a surface reflectance. The atmospheric correction uses water vapor, aerosol, ozone, and Aerosol Optical Thickness values from the Moderate Resolution Imaging Spectroradiometer (MODIS), and digital elevation derived from the Earth Topography Five Minute Grid (ETOPO5) \cite{LandsatCalVal}. The bands taken from each Landsat image are listed in Table \ref{Landsat/Sentinelvartable}.

\begin{table}[ht]
\caption{Specifications of the bands included in joint Landsat 8/9 and Sentinel-1 files.}\label{Landsat/Sentinelvartable}
\begin{tabular}{|p{2.4 cm}|p{1.3 cm}|l|p{2.4 cm}|p{2.4 cm}|}
\hline
\textbf{Band\newline Description} & \textbf{Band Number} & \textbf{Wavelength} & \textbf{Spatial\newline Resolution} & \textbf{Temporal\newline Resolution}\\
\hline
Landsat 8/9 L2\newline blue surface\newline reflectance  & 2  & 0.435-0.451 $\mu$m & 30 m & 16 days each; 8 days for both\\
\hline
Landsat 8/9 L2\newline green surface\newline reflectance & 3  & 0.452-0.512 $\mu$m & 30 m & 16 days each; 8 days for both\\
\hline
Landsat 8/9 L2\newline red surface\newline reflectance   & 4  & 0.636-0.673 $\mu$m & 30 m & 16 days each; 8 days for both\\
\hline
Landsat 8/9 L2\newline NIR surface\newline reflectance   & 5  & 0.851-0.879 $\mu$m & 30 m & 16 days each; 8 days for both\\
\hline
Landsat 8/9 L2\newline SWIR1 surface\newline reflectance & 6  & 1.566-1.651 $\mu$m & 30 m & 16 days each; 8 days for both\\
\hline
Landsat 8/9 L2\newline SWIR2 surface\newline reflectance & 7  & 2.107-2.294 $\mu$m & 30 m & 16 days each; 8 days for both\\
\hline
Landsat 8/9 LST & 10 & 10.60-11.19 $\mu$m & 30 m & 16 days each; 8 days for both\\
\hline
Landsat 8/9\newline Cloud mask & N/A & N/A & 30 m & 16 days each; 8 days for both\\
\hline
Sentinel-1 VV & N/A & 5.55 cm & 10 m (resampled\newline to 30 m) & 12 days (resampled to 8 days)\\
\hline
Sentinel-1 VH & N/A & 5.55 cm & 10 m (resampled\newline to 30 m) & 12 days (resampled to 8 days)\\
\hline
Sentinel-1 HH & N/A & 5.55 cm & 10 m (resampled\newline to 30 m) & 12 days (resampled to 8 days)\\
\hline
Sentinel-1 HV & N/A & 5.55 cm & 10 m (resampled\newline to 30 m) & 12 days (resampled to 8 days)\\
\hline
Sentinel-1\newline incidence angle & N/A & N/A & 10 m (resampled\newline to 30 m) & 12 days (resampled to 8 days)\\
\hline
\end{tabular}
\end{table}

Numerous studies have examined how different land cover types affect surface temperature in urban environments, focusing on green spaces, blue spaces, and impervious surfaces \cite{zhao2020assessing, imran2021impact, hart2009quantifying}. These relationships stem from differences in evapotranspiration, shade, and wind flow among land-cover types, all of which affect local heat transfer with the surrounding environment \cite{zhao2020assessing}. Three commonly used normalized difference indices are the NDVI (band 1=near-infrared [NIR], band 2=red), the Normalized Difference Built-up Index (NDBI; band 1=SWIR1, band 2=NIR), and the Normalized Difference Water Index (NDWI; band 1=green, band 2=NIR).

\begin{equation}\label{NDEquation}
    \text{Normalized difference indices}= \frac{\text{band 1}-\text{band 2}}{\text{band 1}+\text{band 2}}
\end{equation}

\subsection{Sentinel-1}

\subsubsection{Sentinel-1 Overview}

Sentinel-1 consists of active-sensing satellites in a near-polar, sun-synchronous orbit \cite{torres2012gmes}. The first satellite, Sentinel-1A, was launched on 3 April 2014. Sentinel-1B was launched on 25 April 2016 but experienced a power supply problem on 23 December 2021, rendering it unable to deliver data \cite{Sentiwiki}. Sentinel-1C, the replacement for Sentinel-1B, was launched on 5 December 2024, which is outside the temporal coverage of our dataset. A complete orbital cycle for each satellite takes 12 days, with imagery delivered in latitude/longitude coordinates using the WGS-84 coordinate system \cite{Sentiwiki}. Sentinel-1 satellites carry a synthetic aperture radar (SAR) sensor, which helps determine surface roughness, object orientation, and deformation over time. SAR is a satellite-based radar that sends pulses of microwave radiation to the surface and measures the amount scattered back to the sensor. As the satellite orbits, the instrument transmits pulses whose frequencies vary slightly over time, while maintaining a constant amplitude, which enables accurate determination of the ground position of the backscattered signal \cite{Sentiwiki}.

Sentinel-1 SAR has dual-polarization capabilities, which means it can emit and receive both horizontally and vertically polarized radiation. Each Sentinel-1 acquisition in dual-polarization mode includes two polarization channels, consisting of a co-polarization measurement and a cross-polarization measurement, along with metadata describing the local incidence angle for each pixel. The two scene possibilities are [VV, VH, angle] and [HH, HV, angle]. The first letter is the polarization the satellite emits, and the second letter is the polarization the satellite listens for, with ‘V’ for vertical and ‘H’ for horizontal. SAR uses different polarization configurations depending on its location over the Earth, as determined by the selected scanning mode. Sentinel-1 SAR employs four different scanning methods when acquiring data, of which two are used for the images in our dataset. The majority of our Sentinel-1 images are acquired in the Interferometric Wide-Swath (IW) mode at high resolution ($\approx5$ m$\times20$ m). IW is the primary mode for use over land, with VV+VH polarizations \cite{Sentiwiki}. This mode satisfies all mission requirements and builds a consistent database of operations. All other Sentinel-1 images in our dataset are acquired in Stripmap (SM) acquisition mode at high resolution ($\approx5$ m$\times5$ m) \cite{Sentiwiki}. SM is primarily used for small islands and anomalous events. In our dataset, we use the Ground-Range Detected (GRD) product from Sentinel-1 SAR, which provides the amplitude but not the phase of the radar signal.

\subsubsection{Sentinel-1 Variable Selection}

Sentinel-1 SAR transmits at microwave wavelengths ($\approx 5.55$ cm) \cite{Sentiwiki}, allowing surface observations even under cloud cover and thereby complementing Landsat, which passively senses at visible and infrared wavelengths that cannot penetrate clouds. SAR can be used to determine surface roughness, which can help identify rocky surfaces, vegetation, and the boundaries between buildings \cite{koukiou2024sar, zhang2016mapping, sauer2011mapping, gavsparovic2020comparative, bai2021SARLULC, schuler2002surface}. Increasingly large values of backscattered radar signal generally indicate denser vegetation and more densely built urban areas, although this relationship is weaker in very dense or wet vegetation \cite{koukiou2024sar}. Significant differences in backscatter values may indicate a difference in land use between pixels \cite{bai2021SARLULC}. 

We resample the Sentinel-1 SAR GRD from 10 m to 30 m using nearest-neighbor resampling to harmonize with Landsat observations. The dual-polarization capability of Sentinel-1 SAR observations can also help users estimate the shape of objects sensed \cite{zhang2016mapping, sauer2011mapping, koukiou2024sar}. Conceptually, horizontally polarized radiation can be considered to oscillate in the east-west direction, whereas vertically polarized radiation oscillates north-south. Therefore, surfaces exhibiting stronger backscatter in the horizontal than in the vertical polarization are typically oriented along the east-west axis, although this relationship also depends on the acquisition geometry. The bands taken from each Sentinel-1 image are listed in Table \ref{Landsat/Sentinelvartable}. More specifications regarding resampling are available in Section \ref{methodssec}.

\subsection{GOES}

\subsubsection{GOES Overview}

GOES is a series of geostationary satellites operated by NOAA that continuously monitor weather systems and environmental conditions across the Western Hemisphere at high temporal resolution \cite{lindsey2024geoxo}. The current generation (GOES-R series) provides full-disk imagery every 10 minutes, continental U.S. imagery every 5 minutes, and mesoscale imagery as frequently as every 30 seconds. Geostationary satellites orbit the Earth at the same rate as the planet's rotation, and therefore, their position appears fixed relative to the Earth's surface. As a result, they continuously image the same region with each observation.

In our dataset, we use satellite data from the GOES-R series. Two satellites in this geostationary constellation serve the Western Hemisphere: GOES-East (centered at $\approx$ 75$\degree$W) monitors much of the Americas and Atlantic Ocean, and GOES-West (centered at $\approx$ 137$\degree$W) monitors the Pacific basin and western North America (including Hawaii and Alaska). GOES-16 became operational as GOES-East on 18 December 2017 and remained so until it was replaced by GOES-19 on 7 April 2025 as part of routine operations. GOES-17 became GOES-West on 12 February 2019 and was replaced by GOES-18 on 4 January 2023 due to a cooling issue affecting GOES-17's Advanced Baseline Imager (ABI). Given the temporal coverage of our dataset, we utilize data from GOES-16, GOES-17, and GOES-18.

The ABI on board each GOES-R satellite has higher temporal, spatial, and radiometric resolution, as well as improved navigation and registration capabilities, compared to previous GOES generations \cite{schmit2017closer}. The GOES-R ABI has 16 bands: two visible (blue and red), four NIR reflectances, and 10 infrared brightness temperatures. These measurements are projected onto the ABI Fixed Grid. Due to projection distortions when mapping a three-dimensional Earth onto a two-dimensional surface, latitude and longitude coordinates are not suitable to form a grid from the satellite’s viewing perspective. Instead, the grid’s coordinates measure the x and y direction scanning angles in radians \cite{GOEShandbook}. The native spatial resolutions of the ABI are 0.5, 1.0, and 2.0 km at nadir, which correspond to approximately 14, 28, and 56 microradians on the ABI fixed grid \cite{GOEShandbook}.

\subsubsection{GOES Variable Selection}

The high temporal resolution of GOES is useful for monitoring the diurnal variability of urban heat and can complement the infrequent sampling provided by Landsat. More than 90\% of Earth’s emitted radiation comes from longwave- and far-infrared bands (4-100 µm), much of which is absorbed by atmospheric gases (e.g., water vapor and carbon dioxide), which attenuates upwelling longwave radiation before it reaches space \cite{smirnov2019infrared, harries2008far, palchetti2020FIR}. However, relatively little absorption occurs in the atmospheric `window region,' which spans approximately 8 to 12 µm for GOES-R ABI and 8 to 14 µm more broadly \cite{palchetti2020FIR, harries2008far}. Channels within this window are therefore often used to measure LST, while GOES-R ABI Band 16, centered at 13.3 µm, is instead positioned within the CO$_2$ absorption band and used primarily for cloud-top retrievals. As a result, our dataset includes all four GOES ABI brightness temperature bands between 10 and 14 µm (Bands 13-16) to capture the temporal variability of temperature, albeit on a coarser grid (2 km) \cite{palchetti2020FIR} than Landsat (30 m). These four brightness temperature bands are from the Level 2 (L2) cloud and moisture imagery product, which measures top-of-atmosphere (TOA) brightness temperature. A description of these bands, including GOES band numbers and wavelength ranges, is provided in Table \ref{GOES/MW var table}.

\begin{table}[ht]
\caption{Specifications of the bands included in joint GOES/MW LST files.}\label{GOES/MW var table}
\begin{tabular}{|p{2.8 cm}|p{1.2 cm}|l|p{2 cm}|p{2 cm}|}
\hline
\textbf{Band Description} & \textbf{Band\newline Number} & \textbf{Wavelength} & \textbf{Spatial\newline Resolution} & \textbf{Temporal\newline Resolution}\\
\hline
GOES-16/17/18 ABI LWIR BT & 13  & 10.1-10.6 $\mu$m & 2 km & 10 minutes\\
\hline
GOES-16/17/18 ABI LWIR BT & 14  & 10.8-11.6 $\mu$m & 2 km & 10 minutes\\
\hline
GOES-16/17/18 ABI LWIR BT & 15  & 11.8-12.8 $\mu$m & 2 km & 10 minutes\\
\hline
GOES-16/17/18 ABI LWIR BT & 16  & 13.0-13.6 $\mu$m & 2 km & 10 minutes\\
\hline
Microwave LST & N/A  & 0.81-0.83 cm & 0.25 degree \newline(resampled to \newline2 km) & 15 minutes \newline(resampled to \newline10 minutes)\\
\hline
\end{tabular}
\end{table}

\subsection{Microwave (MW) LST} \label{mw_overview}

We include microwave-derived surface temperatures in the dataset, which are effectively cloud-penetrating or `cloud-invariant,' since surface temperature cannot be estimated through clouds using GOES ABI and Landsat infrared bands \cite{prigent2016toward, jones2007satellite}. The wavelength of microwave radiation is sufficiently long that scattering by much smaller cloud particles is negligible and attenuation by rain is relatively small \cite{petty2006first}. This property enables the sensing of near-surface brightness temperatures even under cloudy conditions.

The microwave (MW) LST product comes from microwave radiometers on a constellation of seven low-Earth-orbiting satellites with polar or near-polar inclinations, including both sun-synchronous and non-sun-synchronous orbits: the Advanced Microwave Scanning Radiometer for EOS (AMSR-E) on the Aqua satellite and AMSR2 on the Global Change Observation Mission-Water (GCOM-W1) satellite, the Special Sensor Microwave/Imager (SSM/I) instruments aboard the Defense Meteorological Satellite Program (DMSP) F13, F14, F15, and F16 satellites, the Tropical Rainfall Measuring Mission (TRMM) Microwave Imager (TMI), and the WindSat radiometer on the Coriolis satellite \cite{holmes2015diurnal}. Each radiometer measures vertically polarized Ka-band ($\approx37$ GHz) brightness temperature \cite{holmes2015diurnal}. Specific operational information for each radiometer is presented in Table \ref{mwSat table}. 

\begin{table}[ht]
\caption{Specifications of microwave radiometers providing observations for the MW LST product. Adapted from \protect\citet{holmes2015diurnal}.}\label{mwSat table}
\begin{tabular}{@{}|p{2 cm}|l|l|l|l|@{}}
\hline
\textbf{Sensor} & \textbf{SSM/I} & \textbf{TMI} & \textbf{AMSR2} & \textbf{WindSat} \\
\hline
\textbf{Satellite} & DMSP F13, F14, F15, and F16 & TRMM & GCOM-W & Coriolis \\
\hline
\textbf{Orbit} & Polar & Equatorial & Polar & Polar \\
\hline
\textbf{Equatorial Overpass\newline (Local Time)} & 6-10 AM/PM & Variable & 1:30 AM/PM & 6 AM/PM \\
\hline
\textbf{Operational Accuracy of Ka-band (K)} & 0.4 & 0.5 & 0.7 & 0.5 \\
\hline
\textbf{Spatial Resolution (km)} & 33 & 10 & 12 & 12 \\
\hline
\end{tabular}
\end{table}

Microwave observations are binned onto a 0.25-degree global grid, and the mean of all measurements whose central footprint falls within each new grid cell defines the new brightness temperatures \cite{holmes2015diurnal}. The 0.25-degree grid spacing is determined by the coarsest resolution radiometer (SSM/I), whose horizontal footprint is approximately equivalent at the equator. The MW LST is provided and evaluated at 15-minute intervals to continuously represent surface temperature evolution throughout the day, allowing for comparison with geostationary TIR LST products. We acknowledge that a 0.25-degree grid spacing is too coarse for monitoring intra-city urban heat. Downscaling (i.e., super-resolution) applications are possible; therefore, we include the MW LST product to complement the gaps in infrared-based LST retrievals during cloudy days.

To transform the microwave brightness temperatures into LST, the observations are fitted to a Diurnal Temperature Cycle (DTC) model. The DTC is represented by a harmonic function from the beginning of the day through the early afternoon, followed by an exponential decay during the late-afternoon and nighttime cooling period. The function is described by five parameters: (i) the minimum (morning) temperature, (ii) the maximum (afternoon) temperature, (iii) the time of solar noon, (iv) the time lag between solar noon and the maximum temperature, and (v) a shape parameter controlling the harmonic portion of the function. The temperature parameters are fitted daily for each grid cell, while the time-lag parameter is determined globally across the dataset. The shape and solar-noon parameters are prescribed as functions of the day of year and latitude. The resulting diurnal fit to the Ka-band brightness temperature is converted to LST by calibration to a conventional estimate of LST, typically from TIR sensors \cite{schmetz2002introduction}. In this case, the same DTC formulation was fit to the TIR LST product from the MODIS on Aqua and Terra satellites. The minimum and maximum temperature parameters of each microwave DTC are then scaled by a constant so that their long-term means match those of the TIR-derived LST. The minimum and maximum temperature parameters of each microwave DTC are then scaled by a constant so that their long-term means match those of the TIR-derived LST. More about the microwave LST algorithm and the DTC model can be found in \cite{holmes2018microwave} and \cite{holmes2015diurnal}.

Observations are retained in a grid cell only if they meet four quality-control conditions. First, at least 3 MW and 76 TIR observations must be available in a day to fit the DTC. Second, the minimum MW brightness temperature must be above the freezing point (273 K). Third, the diurnal maximum temperature must occur within one-third of the daylight period from solar noon. Fourth, the root-mean-square error of the MW DTC fit must be either less than 1 K or less than $0.1\bar{6}\times(T_\text{max}-T_\text{min before sunrise})$. Since the DTC formulation assumes smooth daytime warming primarily driven by solar heating in the visible and SWIR spectra, it is only valid on predominantly cloud-free days. The final condition ensures that the DTC shape reflects a clear-sky temperature evolution. Observations that do not meet these requirements and pixels that contain water are set to 0 K.

\section{Methods} \label{methodssec}

\subsection{Location sampling} \label{Sample Locations}

\begin{figure}[h!]
    \centering
    \includegraphics[width=0.9\linewidth]{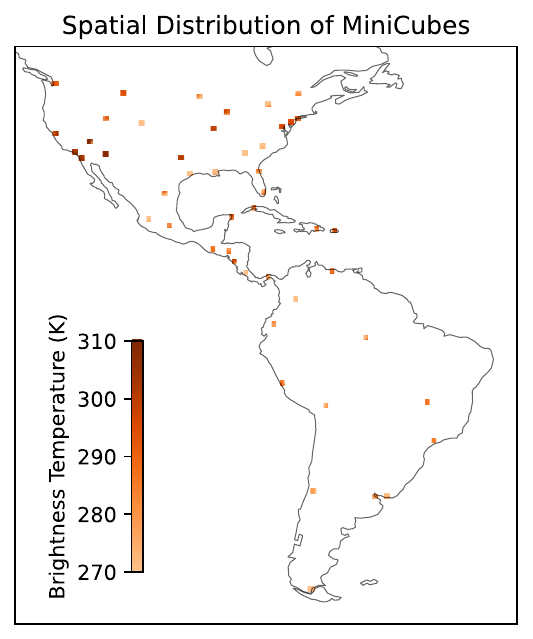}
    \caption{The spatial distribution of all Urban Heat MiniCubes, along with their brightness temperatures on 1 July, 2022.}
    \label{city_map}
\end{figure}

The Urban Heat MiniCubes dataset comprises 48 cities across the Western Hemisphere for the years 2022 and 2023, with each city represented by a 90 km x 90 km grid (Fig. \ref{city_map}). The selected time period provides a recent record of urban heat and enables seasonal analyses, while ensuring that high-resolution observations can be collected from both Landsat 8 and 9. Climatological diversity, total population, and geographic coverage were factors used to select cities for our dataset. Climatological diversity enables the investigation of urban heat across different climate zones. We used the Köppen–Geiger climate classification system to assess climatological representation \cite{peel2007updated, beck2018present}. The designated climate zone for each city is given in Table \ref{city_zones_table}. Higher-population cities concentrate a substantial number of residents and thus a greater potential burden of heat exposure, thereby increasing the dataset's relevance to public health and policy. When numerous high-population cities were geographically clustered, lower population cities in other climate zones were prioritized to prevent over-representation of specific geographic regions in the dataset. We acknowledge that heat-related risks in smaller urban and peri-urban areas may differ systematically from those included in Urban Heat MiniCubes; future work should expand the dataset to include less populated cities. Such an expansion of the dataset would facilitate a more detailed characterization of, and comparison between, different urban heat environments.

\newpage

\subsubsection{Landsat}

Standard Landsat 8/9 multispectral scenes have dimensions of $190\times180$ km, spanning well beyond any single urban area. The nominal pixel resolution of these images is 30 m, but the pixel dimensions vary across images due to Earth's curvature and projection choice. To determine the image dimensions that encompass most cities in our dataset, we based the dimensions on geographically expansive cities, such as Dallas, TX, USA, and Houston, TX, USA. Dimensions of $3000\times3000$ pixels ($\approx90\times90$ km) were deemed sufficient for each `MiniCube' to encompass each chosen city, and we refer to these areas as `export grids.' Larger export grids would include non-urban regions beyond the scope of our dataset, thereby inflating the dataset size and hindering accessibility. The defined export grid is consistent across all satellite-derived data for the respective city.

\begin{figure}[h!]
    \centering
    \includegraphics[width=1\linewidth]{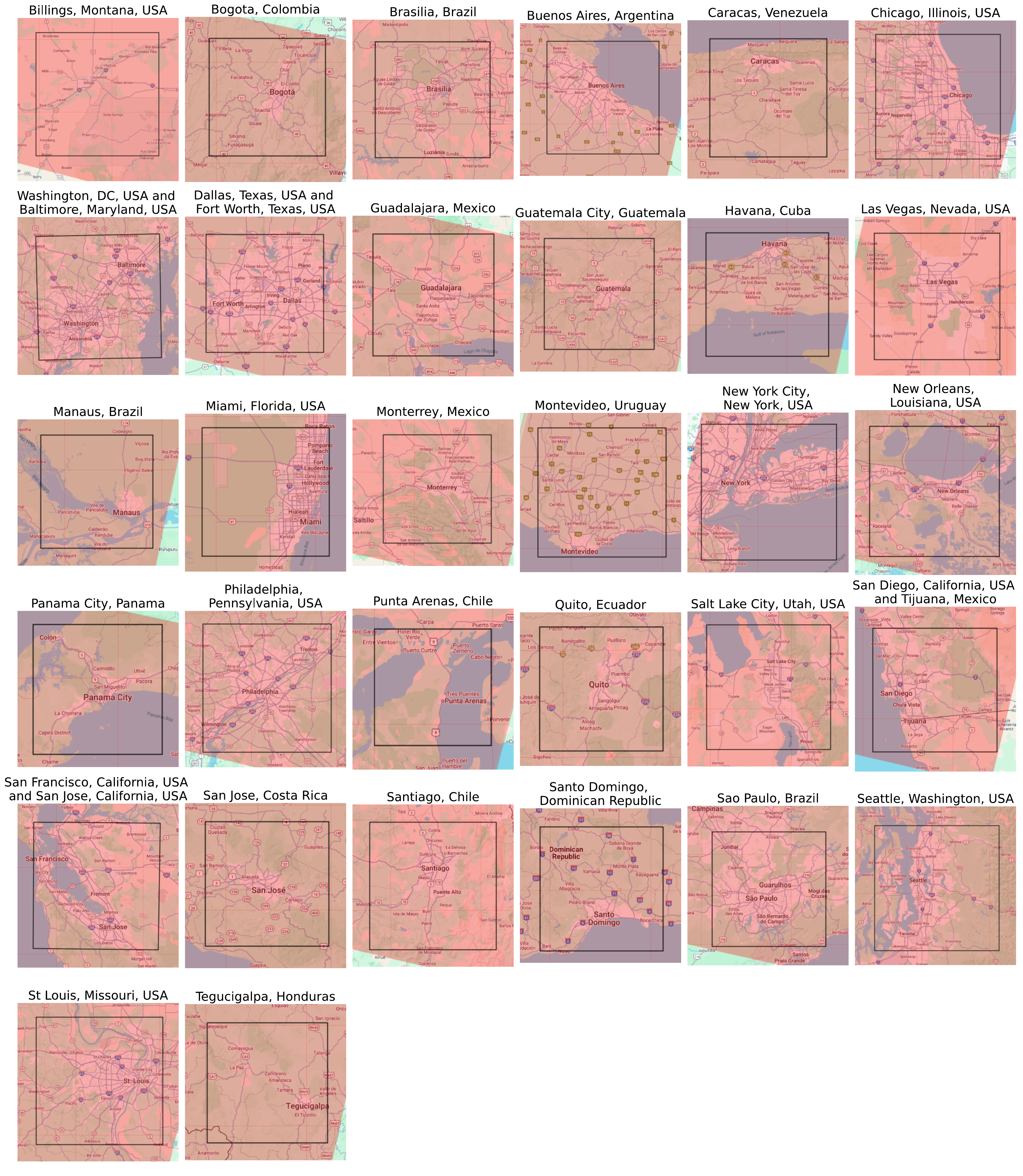}
    \caption{Cities for which one Landsat scene covered the entire export grid (shaded in red). The export grid for each city is shown as a black polygon.}
    \label{one_scene_images}
\end{figure}

Ideally, each export grid would be a $90\times90$ km square centered on the city core. In pixel terms, that corresponds to a grid 3,000 pixels on each side, extending 1,500 pixels (45 km) outward from the city core in every direction. This would allow a single Landsat 8/9 scene acquired during an overpass to cover the entire export grid. Out of 48 cities, more than half met this specified Landsat scene selection criteria (Fig. \ref{one_scene_images}). For other cities, however, factors such as overpass trajectories that only partially covered the export grid, coastal city geometries, and closely adjacent metropolitan areas (e.g., Dallas–Fort Worth, TX, USA) complicated the selection of Landsat 8/9 scenes corresponding to a given export grid and/or the placement of the export grid. Export grids were offset from the urban core for coastal cities with corresponding Landsat overpasses that were farther inland (e.g., San Diego, CA, USA, and Miami, FL, USA) or for cities with two proximal urban cores with a corresponding Landsat overpass (e.g., Washington, D.C., USA, and Baltimore, MD, USA).

\newpage

\begin{figure}[h!]
    \centering
    \includegraphics[width=1\linewidth]{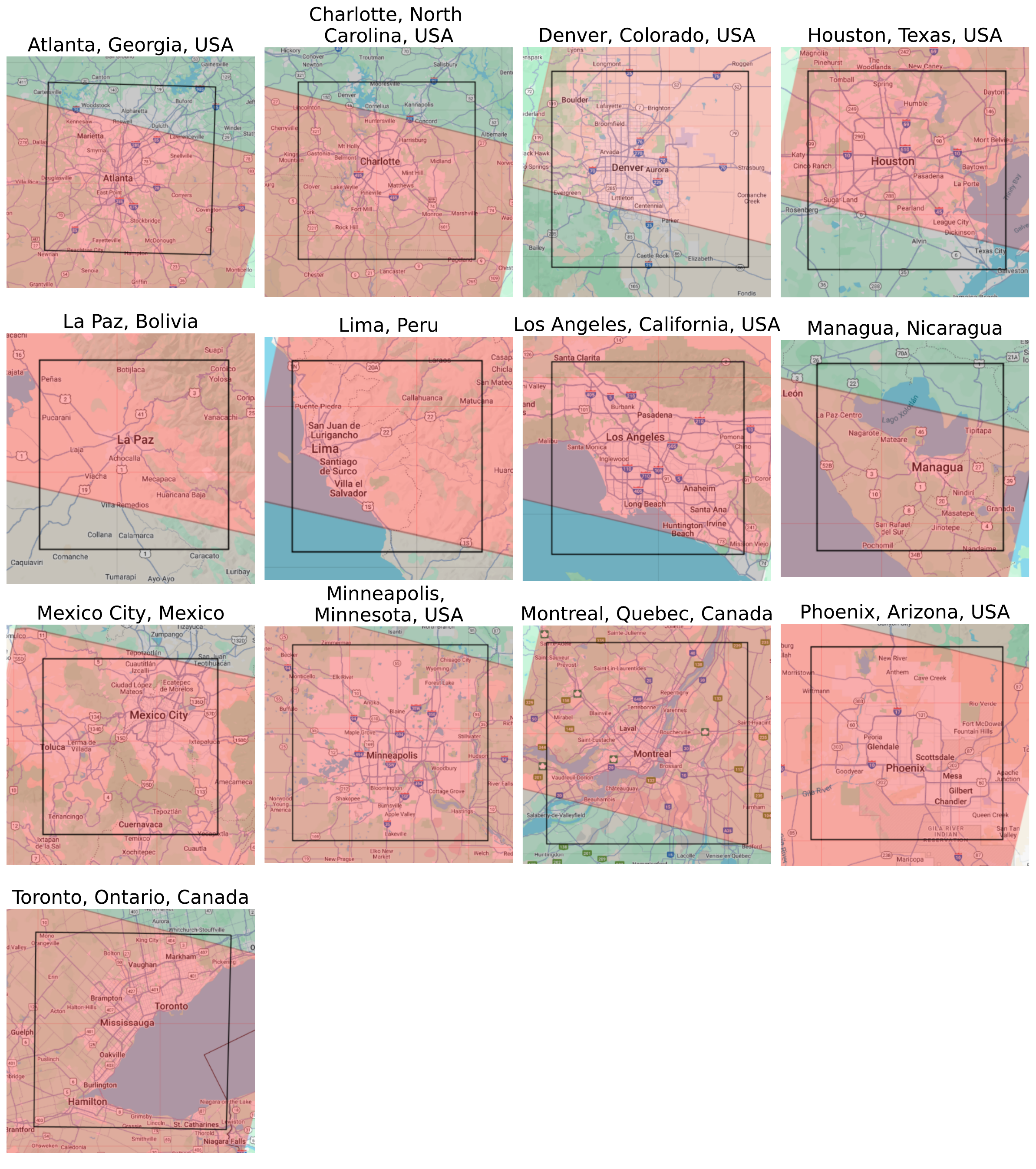}
    \caption{Cities for which two Landsat scenes make up the export grid. The export grid for each city is shown as a black polygon. Adjacent Landsat scenes from an overpass are shaded in red and gray. Landsat scenes shaded in red were mosaicked atop the others because they provided greater coverage of the export grid.}
    \label{two_scene_images}
\end{figure}

For the 15 cities where no Landsat scene entirely covered the export grid, Landsat scenes were mosaicked, with the scene that overlapped most of the export grid being mosaicked on top (Fig. \ref{two_scene_images}). The mosaicking step is motivated by the need to maintain strictly observation-based pixel values. Methods that blend overlapping Landsat scenes (e.g., interpolation or averaging) can create values that are not directly measured by the sensor and may be physically inconsistent. Neighboring Landsat 8/9 scenes acquired within the same orbital pass are collected sequentially and acquired in rapid succession, enabling mosaicking to construct seamless coverage while minimizing temporal inconsistencies.

\begin{figure}[h!]
    \centering
    \includegraphics[width=1\linewidth]{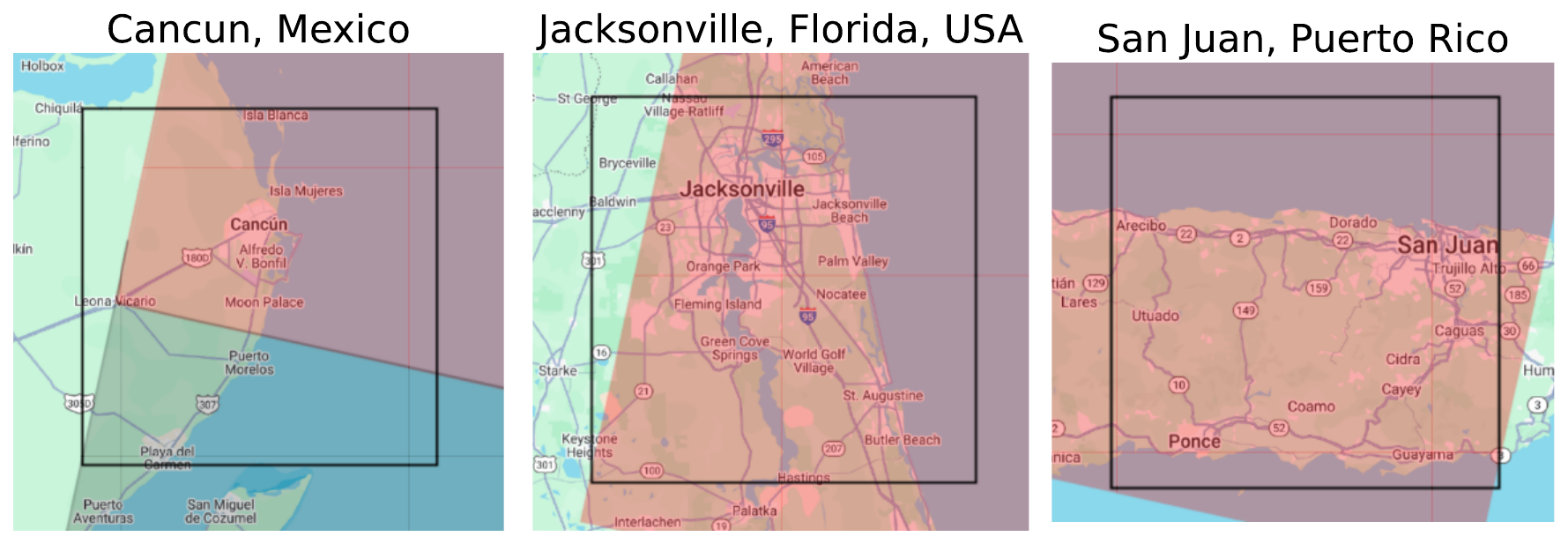}
    \caption{Cities where the full export grid is not covered by Landsat scenes from a single overpass. The export grid for each city is shown as a black polygon. Adjacent Landsat scenes from an overpass are shaded in red and gray. Landsat scenes shaded in red were mosaicked on top because they provided greater coverage of the export grid. Pixels not covered by Landsat scenes are assigned NaNs.}
    \label{complicated_images}
\end{figure}

For select cities, namely Cancun, Mexico, Jacksonville, FL, USA, and San Juan, Puerto Rico, Landsat scenes from a single overpass did not cover the entire export grid (Fig. \ref{complicated_images}). In these cases, the export grids were positioned to maximize coverage of the urban core, even when this criterion resulted in partial scene coverage and missing pixels (e.g., scene boundary gaps or QA-masked pixels). To preserve temporal continuity, each image in the dataset was derived from a single Landsat overpass (i.e., no compositing across multiple overpasses). Missing pixels in any export grid were assigned NaN (Not a Number) values. 


\begin{figure}[ht]
    \centering
    \includegraphics[width=1.0\linewidth]{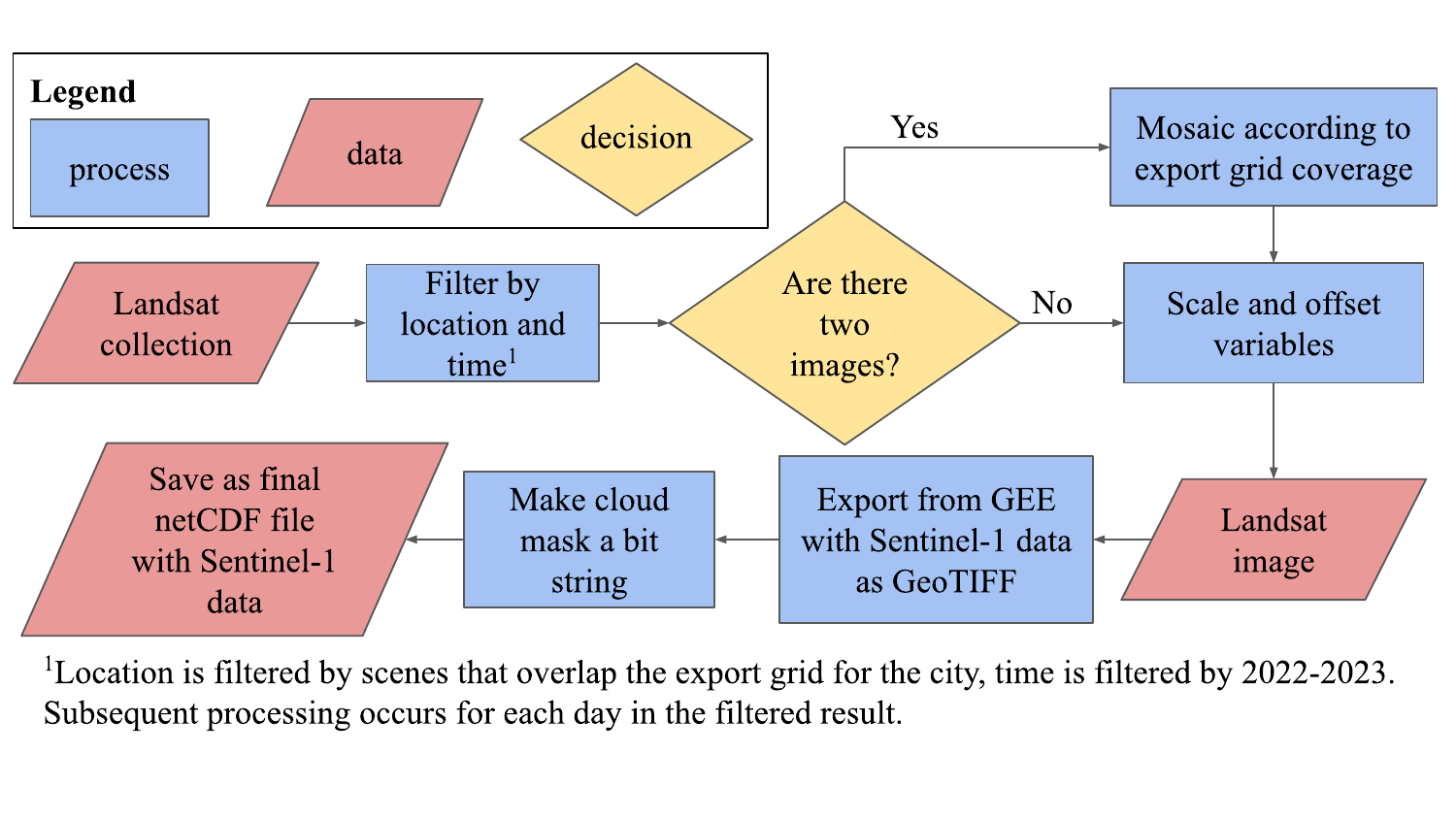}
    \caption{Flowchart describing the Landsat 8/9 data workflow, which starts in the upper-left and ends in the lower-left. Whether the step is a `process,' `data,' or `decision' step is indicated in the legend.}
    \label{Landsat_flowchart}
\end{figure}

After extracting the export grid for a city, the data are multiplied by the corresponding scale terms for each band, then the respective offset terms are added to convert the data to physical units. The download and harmonization with other satellite data are described in Section \ref{geestuff}. After download, we convert the original Landsat cloud mask from base-10 to binary integers and encode it as a binary string. This cloud mask encoding facilitates the interpretation of cloud mask bit values, which contributes to AI-readiness. Users can index the string to retrieve specific bits, and respective bit indices are specified in the file metadata. A Landsat 8/9 workflow overview is provided in Fig. \ref{Landsat_flowchart}.

\newpage

\subsubsection{Sentinel-1}

Despite the native resolution of the Sentinel-1 GRD product being 10 m, we harmonize Sentinel-1 data with Landsat’s spatial and temporal sampling to ensure alignment in space and time. This harmonization is performed because the Landsat-derived 30 m LST product serves as our reference variable for mapping of urban heat. In Google Earth Engine (GEE), the platform used for data download, the Sentinel-1 SAR GRD product is provided in UTM coordinates; therefore, no additional reprojection is performed outside of resampling to the export grid. 

For each city export grid, we select all Sentinel-1 scenes that spatially overlap the grid and are temporally within $\pm6$ days of the Landsat acquisition time. Although Sentinel-1 nominally provides a 6-day revisit interval, Sentinel-1B ceased operation on December 21, 2021, due to an electronics power supply malfunction, and its successor (Sentinel-1C) was not launched until December 5, 2024 \cite{Sentiwiki}. As a result, revisit times during 2022 and 2023 were effectively reduced to 12 days. All selected Sentinel-1 scenes are mosaicked, with observations closest in time to the Landsat acquisition prioritized to minimize temporal mismatch in the final composite. While a 6-day window introduces a potentially substantial temporal offset between Sentinel-1 and Landsat 8/9 measurements, SAR backscatter in urban environments is expected to exhibit limited variability of short (multi-day) intervals given the relatively stable physical structure of the built environment. This temporal tolerance is therefore considered appropriate for constructing a temporally aligned dataset.

\begin{figure}[ht]
    \centering
    \includegraphics[width=1.0\linewidth]{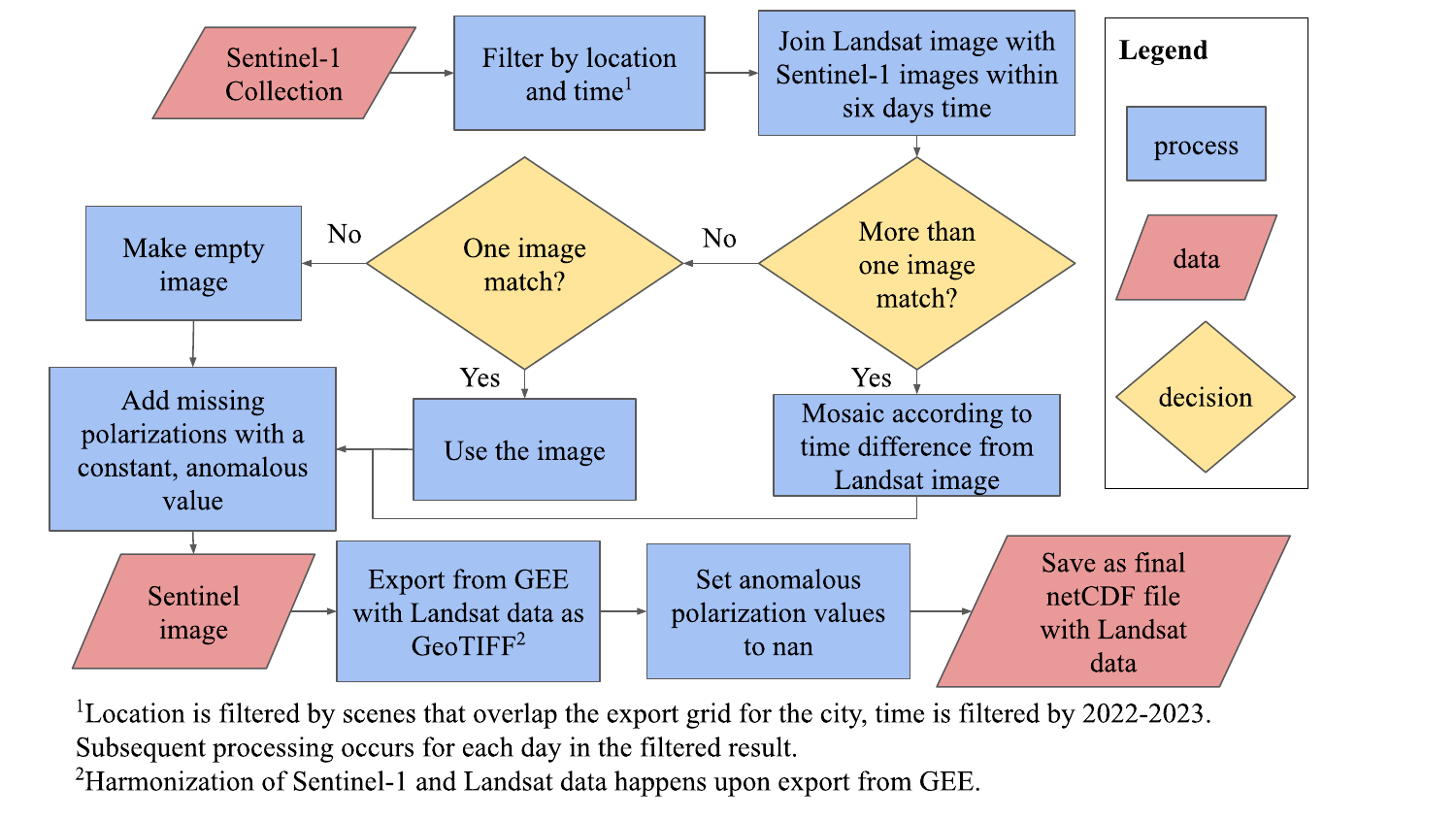}
    \caption{Flowchart describing the Sentinel-1 data workflow, which starts in the upper-left and ends in the lower-right. Whether the step is a `process,' `data,' or `decision' step is indicated in the legend.}
    \label{Sentinel_flowchart}
\end{figure}

No single polarization consistently provides the most recent observation across all export grids; accordingly, all available polarization channels are retained in our dataset. For a given acquisition mode (single- or cross-polarization), pixels not associated with the temporally closest Landsat acquisition are set to NaN. Some images in the dataset lack Sentinel-1 acquisitions within the $\pm6$-day window; in these cases, all polarization channels and the incidence angle layer are entirely NaNs. The Sentinel-1 scanning incidence angle is also retained for subsequent analyses by dataset users. A schematic overview of the processing workflow is shown in Fig. \ref{Sentinel_flowchart}.

\newpage

\subsubsection{GOES}

Each GOES image in the dataset comprises a $45\times45$ pixel subset, obtained by dividing the 90 km image extent by the nominal GOES ABI spatial resolution of 2 km. GOES imagery is provided at 10-minute intervals, corresponding to the native temporal sampling of the GOES full-disk ABI, in order to represent diurnal variability in surface temperature.

Viewing geometry varies across the GOES full-disk domain, with larger off-nadir angles generally associated with increased geometric distortion and potential differences in retrieval performance. To reduce systematic differences attributable to viewing angle, each city is assigned to the GOES satellite that provides a more nadir-proximate observation. The position of GOES-East is $75.2\degree W$ longitude, and GOES-West is $137.0\degree W$ longitude; the midpoint between these longitudes is $106.1\degree W$ longitude, which is used as the primary delineation for assigning cities to GOES-East or GOES-West. All cities follow this delineation except Billings, Montana, USA ($108.5\degree W$), for which GOES-East is used despite its location west of the midpoint. This choice maintains consistency in satellite source for calibration and inter-city comparability, as most cities in the dataset are assigned to GOES-East. The assigned GOES satellite for each city is reported in Table \ref{city_zones_table}.

\begin{figure}[ht]
    \centering
    \includegraphics[width=1.0\linewidth]{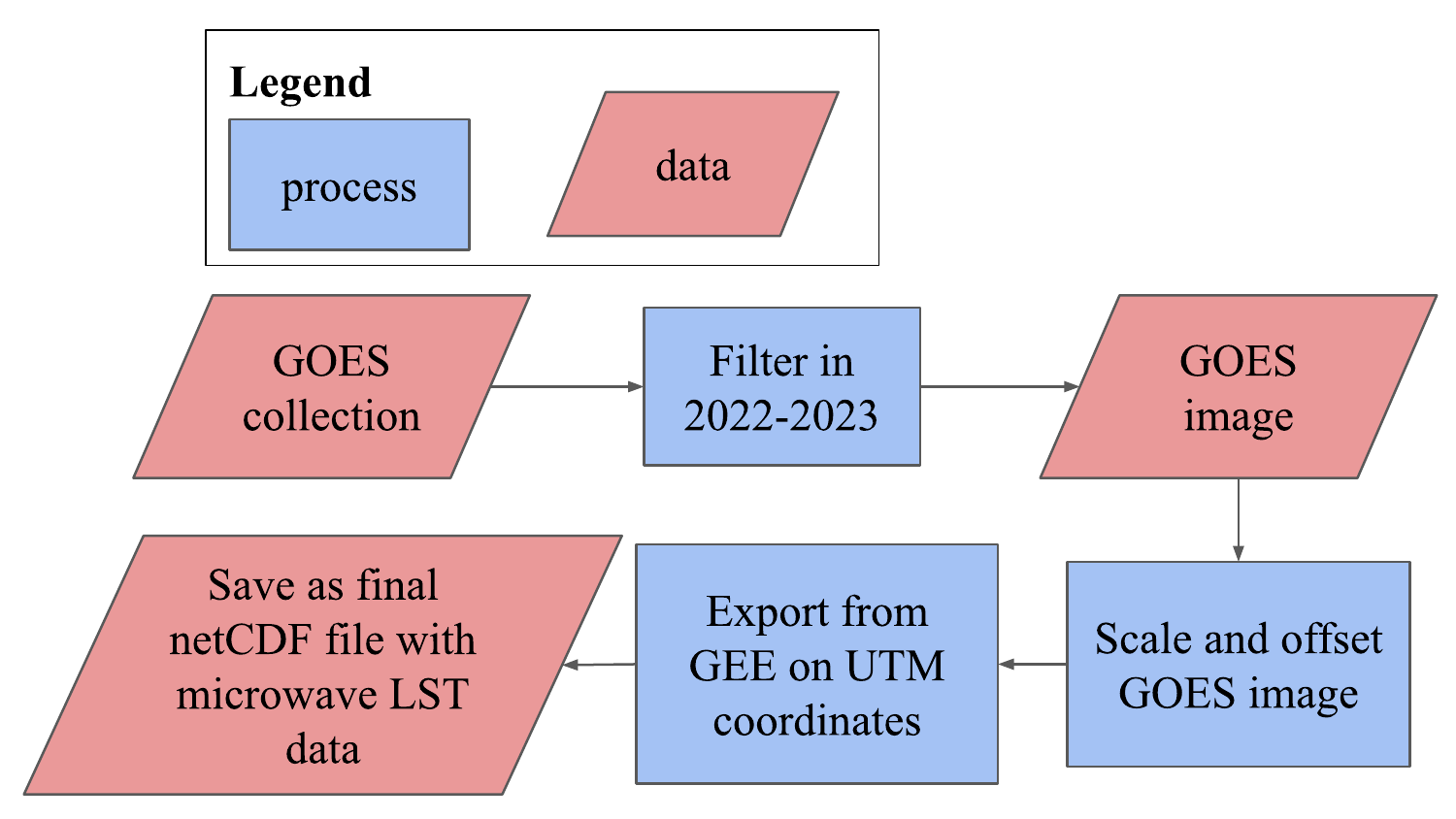}
    \caption{Flowchart describing the GOES data workflow, which starts in the upper-left and ends in the lower-left. Whether the step is a `process' or `data' step is indicated in the legend.}
    \label{GOES_flowchart}
\end{figure}

During processing, digital numbers are converted to physical units by applying band-specific scale factors followed by corresponding offsets. The resulting products are then reprojected to UTM coordinates. An overview of these workflow steps is given in Fig. \ref{GOES_flowchart}.

\subsubsection{Microwave LST}

The microwave LST product is provided at 15-minute intervals. Microwave LST fields are interpolated onto the GOES spatial grid to facilitate pixel-wise comparisons and joint analysis at comparable temporal sampling rates. Temporal alignment is determined using time-adjusted microwave LST timestamps reported in local solar time. UTC time is computed from local solar time and longitude according to the equation below.
\begin{equation}
    \text{UTC time} = \text{Local Solar Time} - \left(\frac{\text{Longitude}}{360}\right) \times 24
    \label{local solar time}
\end{equation}

\noindent This procedure yields a unique microwave timestamp for each observation, which is rounded to the nearest 10-minute increment for interpolation to GOES imagery.

\begin{figure}[ht]
    \centering
    \includegraphics[width=1.0\linewidth]{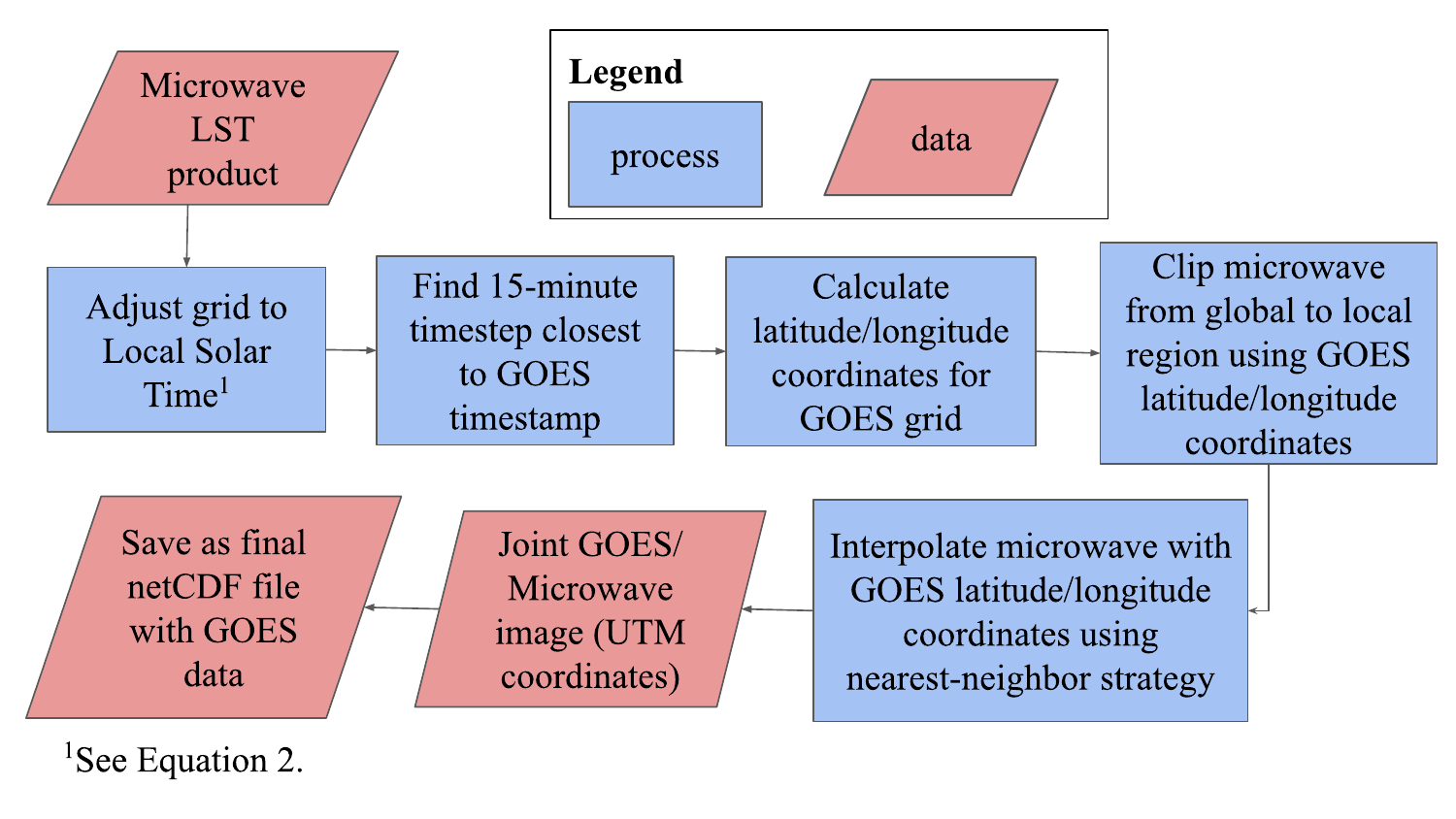}
    \caption{Flowchart describing the microwave LST data workflow, which starts in the upper-left and ends in the lower-left. Whether the step is a `process' or `data' step is indicated in the legend.}
    \label{mw_LST_flowchart}
\end{figure}

The microwave product has substantially coarser spatial resolution than GOES (0.25 degrees versus 2 km). When resampling the microwave LST fields onto the GOES grid, we apply nearest-neighbor interpolation to preserve the original microwave retrieval values and avoid smoothing artifacts that can arise from higher-order interpolation. The conservative resampling choice also enables downstream users to apply alternative aggregation or interpolation strategies as needed for specific analyses. The order of these workflow steps is given in Fig. \ref{mw_LST_flowchart}.

\newpage

\subsection{AI-Ready Data Generation}\label{geestuff}

We used GEE \cite{gorelick2017google} to source, partially preprocess, and spatially align the remote sensing data on a common analysis grid. GEE is free for verified noncommercial use (e.g., academic and research applications), while commercial or operational use is provided through paid Google Cloud offerings. The underlying Landsat 8/9, GOES-R, and Sentinel-1 datasets are also publicly available outside of GEE, as described in Section \ref{DataAvailavility}. 

All variables in the dataset are projected to UTM coordinates by resampling source imagery onto an export grid defined for each urban area. Each export grid is constructed in UTM coordinates such that the easting and northing of the grid corners are integer multiples of the target spatial resolution (30 m for Landsat 8/9 and 2 km for GOES-R). Grid cell centers are then generated at uniform increments of the target resolution within the corresponding UTM zone. Pixel values are assigned to the export grid using nearest-neighbor resampling. For source imagery provided in an alternative projection, source pixel locations are first transformed into UTM coordinates, and the nearest export-grid cell is selected accordingly. Nearest-neighbor resampling preserves original pixel values and minimizes spatial smoothing, which is particularly important for analyses of fine-scale thermal heterogeneity in urban environments. Separate coordinate files providing the corresponding latitude and longitude are provided for both data modalities. These coordinates are computed by converting each UTM grid location to geographic (latitude and longitude) coordinates.

UTM coordinates are calculated by projecting geographic coordinates (latitude and longitude) into one of the sixty UTM zones, each spanning six degrees of longitude, using a transverse Mercator projection \cite{moore1997transverse}. The central meridian of each zone is assigned a scale factor of 0.9996 to reduce projection distortion \cite{brooks1973universal}. A false easting value of 500 km is applied at the central meridian, such that locations east of the meridian have larger easting values, whereas locations west of the meridian have smaller easting values. In the Northern Hemisphere, northing is defined to increase from 0 km at the equator, whereas in the Southern Hemisphere, a false northing of 10,000 km is applied at the equator with values decreasing southward \cite{brooks1973universal}. The inverse transformation is used to derive latitude and longitude coordinates from UTM coordinates.

\section{Data Records}

\subsection{Storage}

\begin{figure}[h!]
    \centering
    \includegraphics[width=1.0\linewidth]{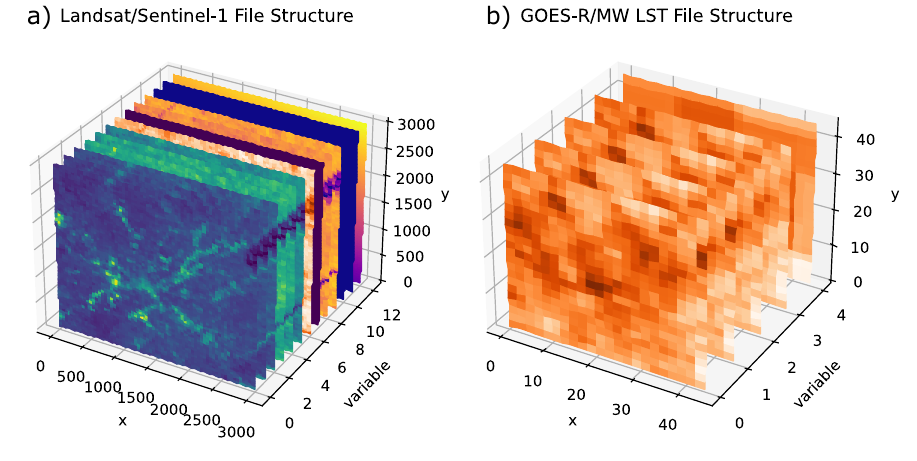}
    \caption{Illustration of the dataset file structure, taken from 16:10 UTC on 4 January 2022 over Atlanta, Georgia, USA. a) The structure of Landsat8/9 and Sentinel-1 data files, including 13 variables, each with dimensions of $3000\times3000$ pixels. See Table \ref{Landsat/Sentinelvartable} for more details. b) The structure of GOES-R and MW LST files, including 5 variables, each with dimensions of $45\times45$ pixels. See Table \ref{GOES/MW var table} for more details.}
   \label{data structure fig}
\end{figure}

Urban Heat MiniCubes consists of two data file types. The first contains high-spatial-resolution observations from Landsat 8/9 and Sentinel-1, acquired every 8 days. Over the dataset's 730-day span (2022-2023), the maximum number of files of this type per city is 92, assuming no Landsat observations are missing. These images contain 13 bands in total. Six bands are L2 Landsat 8/9 surface reflectance products, including red, green, and blue (RGB), one NIR, and two SWIR bands. Additional Landsat bands include the L2 Landsat LST product and a cloud mask. Four bands are Sentinel-1 SAR backscatter values corresponding to the VV, VH, HH, and HV polarizations, and the final band is the Sentinel-1 incidence angle. Full descriptions of these bands, including Landsat band numbers and wavelength ranges, are provided in Table \ref{Landsat/Sentinelvartable}. The structure of this file type is illustrated in Fig. \ref{data structure fig}a.

The second satellite file type contains low-spatial-resolution observations acquired every 10 minutes from GOES-16, GOES-17, or GOES-18, along with a microwave LST product derived from diurnal temperature cycles. For each city using GOES-East observations, there are 104730 observations of this type, whereas for each city using GOES-West observations, there are 103146 observations. Files of this type are concatenated along the time dimension, resulting in a single data file per city to reduce the overall number of files and facilitate data transfer/sharing. These images contain five bands. Four bands are L2 GOES longwave infrared (LWIR) brightness temperatures (BT), and the remaining band is the microwave LST. The full description of these bands, including GOES band numbers and wavelength ranges, is provided in Table \ref{GOES/MW var table}. The structure of this file type is pictured in Fig. \ref{data structure fig}b.

\begin{figure}[h!]
    \centering
    \includegraphics[width=1.0\linewidth]{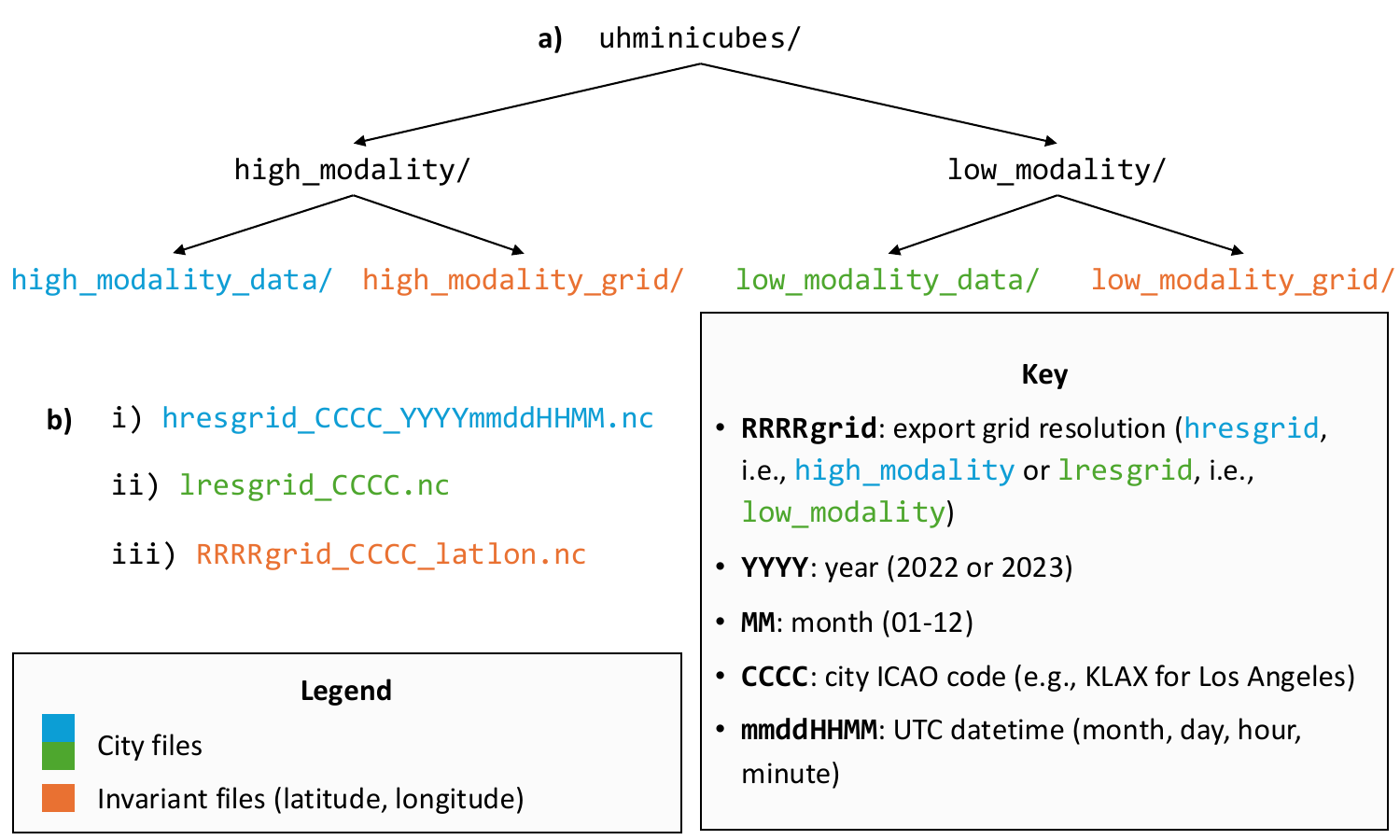}
    \caption{Illustration of the dataset directory structure, with terms as defined by the key. The ICAO (International Civil Aviation Organization) code (i.e., CCCC) for each city in the dataset is provided in Table \ref{city_zones_table}. a) The dataset directory tree, with the top being the root directory. Leaf directories that contain Landsat/Sentinel-1 or GOES-R/MW files are colored in blue and green, respectively, and leaf directories that contain latitude and longitude coordinates are colored in orange, as indicated in the legend. b) Format of file names in the dataset for (i, blue) Landsat/Sentinel-1, (ii, green) GOES-R/MW LST, and (iii, orange) latitude and longitude coordinate files.}
   \label{directory_tree_fig}
\end{figure}

Both file types are provided in netCDF format with UTM coordinates. Separate netCDF files containing the corresponding latitude and longitude coordinate arrays for both file types are also provided. The directory tree structure showing how files are stored in the dataset and the general filename format of each file is given in Fig. \ref{directory_tree_fig}.

\subsection{Metadata}

Data in Urban Heat MiniCubes follows the NetCDF Climate and Forecast (CF) metadata conventions \cite{CF_conventions}. The CF conventions are designed to promote easy processing and interpretation of NetCDF files. Each variable is assigned a definitive description, known as a "standard name,'' enabling straightforward interpretation of dataset contents. Coordinate metadata details the spatial and temporal properties of the data, including the coordinate system and the datetime calendar used. Physical variable metadata includes units, missing values, wavelength, and valid min and max values, when appropriate. Descriptions of metadata tags present in the dataset are provided in Table  \ref{metadata_table}.

\begin{table}[ht]
\caption{Metadata tags present in dataset files. ``Metadata Location'' describes the location where the metadata is included; the three distinctions are: ``file'' (metadata stored in the file attributes), ``coordinates'' (metadata stored in coordinate level), and ``variables'' (metadata stored in variable level).}\label{metadata_table}
\begin{tabular}{|l|p{2 cm}|p{1 cm}|p{6 cm}|}
\hline
\textbf{Metadata Tag} & \textbf{Metadata\newline Location} & \textbf{dtype} & \textbf{Interpretation}\\
\hline
title & file & string & Provides a general description of file contents.\\
\hline
institution & file & string & Institution where the data was produced.\\
\hline
source & file & string & Method of production for the original data.\\
\hline
standard\_name & coordinates,\newline variables & string & Definitive identifier of the variable's contents.\\
\hline
long\_name & coordinates,\newline variables & string & Descriptive identifier of the variable's contents.\\
\hline
units & coordinates,\newline variables & string & Units for interpreting the variable's data.\\
\hline
calendar & coordinates & string & Set of datetimes used for a time-based coordinate.\\
\hline
valid\_min & variables & float & Minimum value that constitutes a valid observation for a given variable.\\
\hline
valid\_max & variables & float & Maximum value that constitutes a valid observation for a given variable.\\
\hline
missing\_value & variables & float & Value that indicates a missing observation in a variable's data.\\
\hline
wavelength & variables & string & Wavelength range for a variable's observations.\\
\hline
grid\_mapping & variables & string & Indicates the variable that contains information about the coordinate reference system used for the projection of each variable.\\
\hline
bitmask\_key & variables & string & Key for interpreting the values of the Landsat cloud mask.\\
\hline
\end{tabular}
\end{table}

\newpage

\section{Technical Validation}

\subsection{Validation} \label{validation_section}

Here, we discuss two types of accuracy for the satellites included in the dataset: cartographic accuracy and measurement accuracy. Cartographic accuracy quantifies how closely a measurement matches its true geographic location. When distinguished, pointing accuracy is the degree to which the satellite's reported orientation matches its true orientation, whereas attitude accuracy is the degree to which the satellite can reorient itself. Measurement accuracy describes errors in physically measured quantities, including absolute accuracy, the potential error in any individual measurement, and repeatability accuracy, the potential error when the same measurement is repeated under identical conditions.

The maximum cartographic error for all processed Landsat 8/9 L2 products is 12 m for all variables \cite{LandsatHandbook}. Temperatures derived from TIRS bands 10 and 11 have uncertainties of 1 K and 2 K (1$\sigma$), respectively \cite{LandsatHandbook}. OLI TOA reflectances are calibrated to within 3\% radiometric uncertainty \cite{LandsatHandbook}. Landsat 8/9 signal-to-noise ratio and noise-equivalent delta temperature values are provided in Tables 2-2 and 2-3 of the product user guide \cite{LandsatHandbook}.

The maximum cartographic error of the GOES L2 Cloud and Moisture product included in Urban Heat MiniCubes is 1 km for all ABI bands \cite{GOEShandbook}, whereas the measurement accuracy of GOES L2 products is not specified therein. However, the Level-1 (L1) brightness temperatures (not included in Urban Heat MiniCubes) have a maximum repeatability error of 0.2 K and a maximum absolute error of 1.0 K \cite{GOESperformance}.

The maximum pointing error for Sentinel-1 is 0.004$\degree$ for each axis (pitch, roll, and yaw), with a maximum attitude error of approximately 0.01$\degree$ for each axis \cite{Sentiwiki}. The majority of Sentinel-1 images in Urban Heat MiniCubes were acquired in interferometric wide-swath (IW) mode at high resolution, which has a maximum absolute location error of 7 m. All other Sentinel-1 images were acquired in stripmap (SM) mode at high resolution, which has a maximum absolute location error of 2.5 m \cite{Sentiwiki}. The maximum radiometric (absolute) error for all acquisition modes is 1 dB, and the noise-equivalent sigma naught for Sentinel-1 is $-$22 dB \cite{Sentiwiki}.

The European Space Agency (ESA) and the Sentinel-1 Mission Performance Centre, responsible for Sentinel-1 instrument maintenance and processing algorithm updates \cite{Sentiwiki}, do not provide an explicit per-pixel quality flag or reliability metric for Sentinel-1 SAR products. However, auxiliary variables, such as the local incidence angle, can help users assess pixel-level reliability for applications such as classification and change detection. Users are encouraged to consult ESA's annual performance and data quality reports \cite{Sentiwiki}, which summarize key quality indicators, known issues, and overall mission stability. The processing algorithms implemented in the Copernicus Space Component Ground Segment are designed to ensure that Sentinel-1 operational products meet radiometric and geometric quality specifications \cite{Sentiwiki}.

Nearest-neighbor resampling is used when projecting data onto a new grid. The maximum cartographic error occurs when an observation lies at a pixel corner in the reprojected grid, corresponding to a distance of $\sqrt{2}$ times half the pixel resolution. Thus, the maximum cartographic error associated with resampling is 1414 m for GOES, 21.2 m for Landsat, and 7.07 m for Sentinel-1. These errors are additive to the cartographic uncertainties already present in the source datasets, and the resulting misalignments may occur throughout the dataset.

\subsection{Data Analysis: Statistical Overview}

This section presents diagnostic analyses to characterize the statistical structure of Urban Heat MiniCubes and to quantify systematic differences among sensors relevant to surface- and near-surface thermal monitoring. Urban Heat MiniCubes integrates observations from instruments with substantially different spatial resolutions, retrieval algorithms, and measurement definitions (e.g., Landsat LST versus GOES BT), which makes direct cross-sensor comparisons nontrivial. To facilitate comparison while minimizing temporal mismatch, we use the harmonized image products and pair each Landsat acquisition with the temporally closest GOES-R ABI Band 14 BT observation. GOES ABI Band 14 ($11.2\mu$m) is the canonical longwave ``window'' channel and is commonly used as an input in many GOES LST algorithms \cite{GOESLSTalgorithm}. GOES ABI Band 14 provides the most straightforward single-channel thermal reference from the ABI for comparison with Landsat LST, and we therefore use it for a first-order diagnostic comparison, though it measures top-of-the-atmosphere BT and is not an exact comparison to LST.

\begin{figure}[htbp!]
    \centering
    \includegraphics[width=0.95\linewidth]{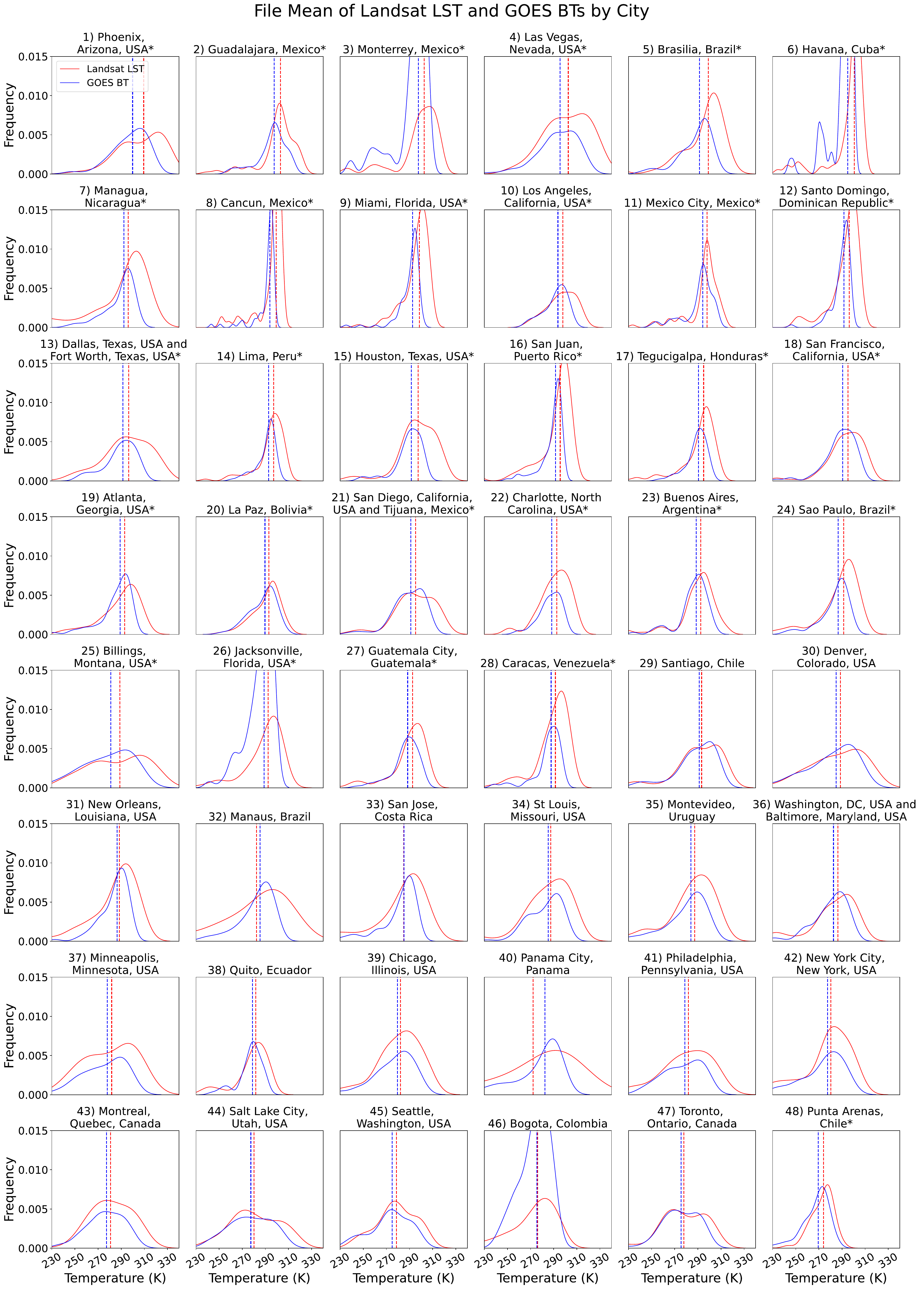}
    \caption{PDFs of Landsat LST and GOES Band 14 BT for each city in the dataset. For each Landsat image, GOES data are taken from the temporally closest GOES observation. PDFs are constructed from the acquisition-average temperature values for a given city. Asterisks next to city names indicate statistically significant differences ($p<0.05$) between Landsat LST and GOES BT distributions based on a Mann-Whitney U test (two-tailed). Dashed vertical lines are added to the midpoint of each distribution (Cumulative Distribution Function (CDF)$=$0.5) to more easily visualize differences between satellites. Cities are ordered by the median value of the file-averaged Landsat LST, from warmest (1) to coolest (48).}
    \label{temp_comparisons}
\end{figure}

We first assess differences in value distributions between acquisition-averaged Landsat LST and GOES-R ABI Band 14 BT by estimating and comparing probability density functions (PDFs; Fig. \ref{temp_comparisons}). PDFs were generated using univariate kernel density estimation (KDE) \cite{wkeglarczyk2018kernel, chen2017tutorial}, which provides a smooth estimate of the underlying probability density and is less sensitive to binning choices than histograms. In this approach, a Gaussian kernel with a predefined bandwidth is centered at each sample, and the kernels are summed and normalized such that the total area under the curve equals one. Bandwidth is selected using Scott's rule of thumb (Equation \ref{bandwidth equation}), a standard choice for approximately Gaussian distributions \cite{scott1979optimal}.

\begin{equation}
    \text{bandwidth} = 1.059\times \text{A}\times n^{-0.2},
    \label{bandwidth equation}
\end{equation}

\noindent where

\begin{equation}
    \text{A}=\text{min}\left(\text{std}(x), \frac{\text{IQR}(x)}{1.34}\right),
\end{equation}

\noindent and where

\begin{equation}
    \text{std}(x)=\sqrt{\frac{\sum_{i=1}^{n}(x_i-\bar{x})^2}{n-1}},
\end{equation}

\noindent given that IQR is the interquartile range and $n$ the number of observed values $x$. To evaluate whether distributions differ significantly, we apply the Mann-Whitney U test, which tests the null hypothesis that two samples are drawn from the same underlying distribution \cite{mcknight2010mann}. This test is nonparametric and therefore does not require an assumption of normality.

Figure \ref{temp_comparisons} indicates that Landsat LST is warmer on average than GOES BT Band 14 for the majority of cities in the dataset, with weaker or non-significant differences in cooler cities (Mann-Whitney U; no asterisk). This result is expected, as Landsat LST is a surface retrieval that includes atmospheric compensation, whereas GOES Band 14 is a top-of-the-atmosphere BT measurement. Importantly, we did not explicitly filter pixels with cloud coverage, so GOES BT distributions may include cold cloud-top temperatures. Consequently, the observed offsets should be interpreted as a diagnostic inter-sensor difference under mixed-sky conditions, not as an equivalence or validation between Landsat LST and GOES BT measurements.

\begin{figure}[htbp!]
    \centering
    \includegraphics[width=0.95\linewidth]{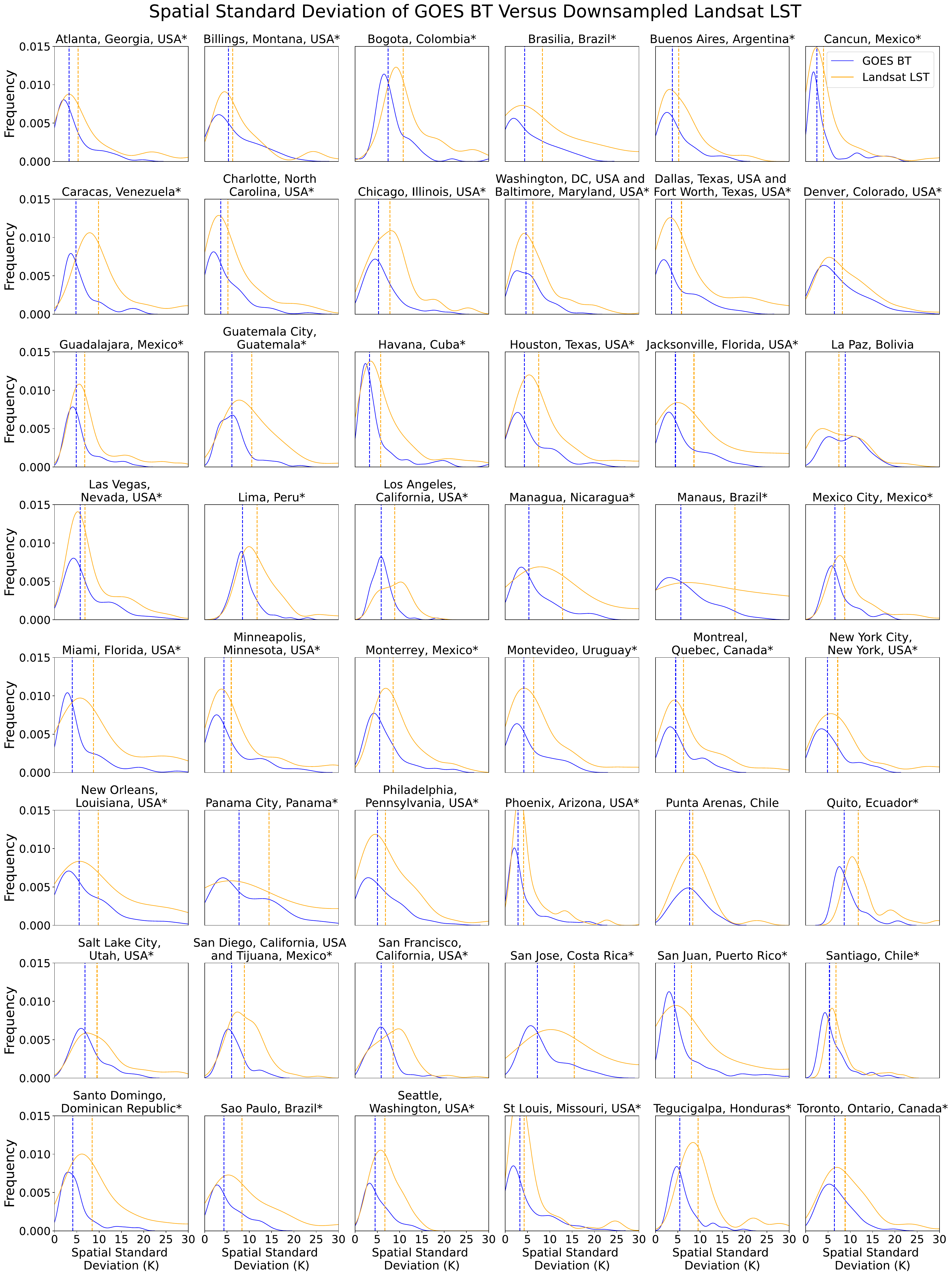}
    \caption{PDFs of within-image spatial standard deviation for GOES ABI Band 14 BT and filtered/resampled Landsat LST for each city in the dataset. Landsat LST images are first smoothed with a 2D Gaussian filter and then resampled to the coarser GOES image using bicubic resampling. For each Landsat acquisition, GOES data are taken from the temporally closest GOES observation. Asterisks next to city names indicate the Landsat LST distribution to be statistically significantly greater than the GOES BT distribution ($p < 0.05$) based on a Mann-Whitney U test (one-tailed, Landsat LST $>$ GOES Band 14 BT). Dashed vertical lines are added to the midpoint of each distribution (CDF$=$0.5) to more easily visualize differences between satellites.}
    \label{GOES filtered Landsat std variability}
\end{figure}

Landsat LST exhibits a larger within-image spatial variability than GOES ABI Band 14 BT (seen in Figs. \ref{LST std seasonal variability} and \ref{GOES BT std seasonal variability}), consistent with Landsat's finer spatial resolution and its ability to resolve sub-pixel thermal heterogeneity that is spatially averaged in GOES ABI observations. To evaluate whether these differences are primarily attributable to spatial resolution, we perform a scale-matching analysis in which Landsat LST is low-pass filtered and resampled to the coarser GOES grid. Specifically, we apply a 2D Gaussian low-pass filter to each Landsat LST image, then resample (coarsen) the filtered field onto the GOES pixel grid using bicubic interpolation. We then compute the spatial standard deviation of pixel values within each resulting $45\times45$ grid for every matched acquisition and compare the resulting PDFs of these within-image spatial standard deviations between GOES Band 14 BT and the filtered/resampled Landsat LST (Fig. \ref{GOES filtered Landsat std variability}). For each city, a one-tailed Mann-Whitney U test evaluates the alternative hypothesis that the filtered/resampled Landsat LST retains greater within-image spatial variability than GOES Band 14 BT. Across all cities in the dataset, except La Paz, Bolivia, and Punta Arenas, Chile, the filtered/resampled Landsat LST exhibits significantly higher spatial variability ($p < 0.05$), suggesting that the observed variability differences are not fully removed by this approximate scale matching and likely reflect a combination of spatial support and measurement-definition effects. We also note that within-image spatial variability (Figs. \ref{LST std seasonal variability}-\ref{GOES BT std seasonal variability}) is generally greater during boreal (JJA) and austral (DJF) summer for both Landsat LST and GOES Band 14 BT (35 of 47 cities, and 23 of 47, respectively) and less so during boreal (DJF) and austral (JJA) winter (40 of 47 cities, and 25 of 47, respectively). Quito, Ecuador, was excluded from this analysis due to its equatorial location and weak annual temperature seasonality.

PDFs showing the seasonal variability of Landsat LST and GOES Band 14 BT are provided in Fig. \ref{LST/BT seasonal variability}, along with vertical lines to show which season is the warmest and coldest. Across the 48 cities, Landsat LST and GOES BT agree on the season of peak cooling in 44 cases and on the season of peak warming in 37 cases. Where the timing differs, mismatches occur predominantly in tropical cities, consistent with their weaker seasonal temperature amplitude, but they also appear in several extratropical cities (Monterrey, New York City, and Phoenix) for peak warming only. These mismatch cases should be treated with particular caution in any Landsat-to-GOES mappings, because the inferred timing of peak warming/cooling can be sensor-dependent. In terms of magnitude, Landsat LST is warmer than GOES BT in 44 cities at peak cooling and in all 48 cities at peak warming, indicating a systematic offset in which Landsat LST exceeds GOES Band 14 BT, consistent with Fig. \ref{temp_comparisons}. 

City-specific PDFs of Landsat cloud coverage (percentage of cloudy pixels per acquisition) are provided in Fig. \ref{Cloud cover pdfs}, ordered from most (Bogota, Colombia) to least (Phoenix, Arizona, USA) cloudy city. Overall, cloud cover is higher in cities in tropical Köppen–Geiger climate zones \cite{peel2007updated, beck2018present}, likely related to moisture availability and convection, and lower in desert and arid/semi-arid regions, where precipitation is suppressed, all of which are important considerations for dataset use.

\begin{figure}
    \centering
    \includegraphics[width=1.0\linewidth]{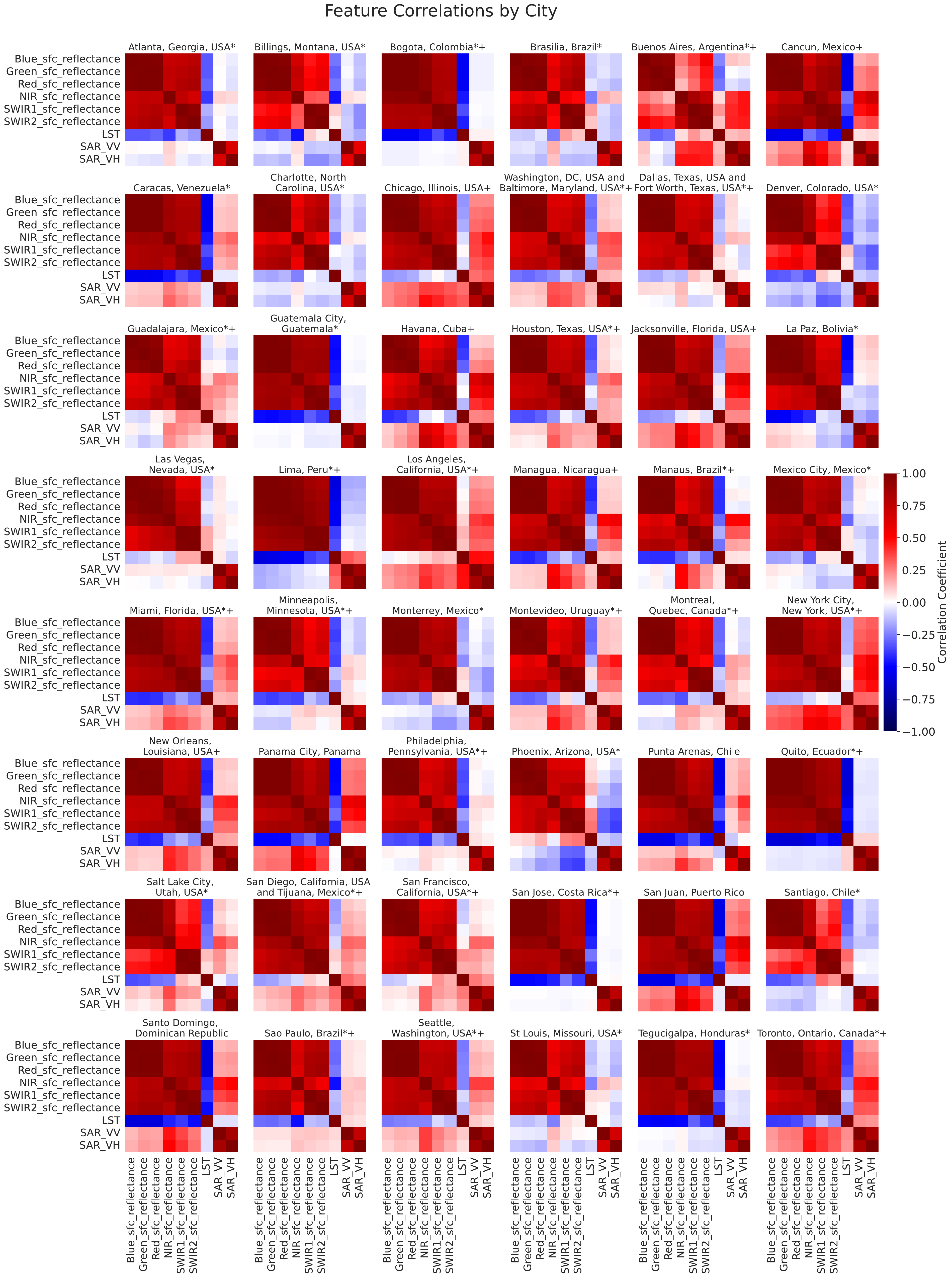}
    \caption{Heatmaps showing the Pearson correlation coefficients between feature variables in Landsat/Sentinel-1 files for each city in the dataset. Asterisks next to city names indicate that SWIR1 and SWIR2 have the largest correlations with Landsat LST out of all of the surface reflectances. Plus signs next to city names indicate that SAR VV and VH have positive correlations with Landsat LST.}
    \label{Var corrs}
\end{figure}

Heatmaps summarizing the mean Pearson correlation coefficients among variables in the Landsat/Sentinel-1 files for each city are shown in Fig. \ref{Var corrs}. Landsat LST is retrieved from LWIR radiance, so we evaluate whether the longest wavelength Landsat surface reflectance bands available in our feature set tend to exhibit comparatively stronger positive correlations with LST than shorter-wavelength reflectances. This pattern is observed in 38 of 48 cities in the dataset (asterisks in Fig. \ref{Var corrs}). The sign of the correlation between LST and Sentinel-1 SAR backscatter (VV and VH) is also assessed. In urban environments, higher backscatter can be associated with increased surface roughness and building density, since man-made structures often produce strong radar returns \cite{koukiou2024sar}. However, the LST-backscatter relationship is not necessarily monotonic across land-cover and climate contexts. Smooth, bare, and dry surfaces (e.g., deserts) can exhibit relatively low backscatter while also exhibiting elevated LST, thereby weakening or reversing city-level correlations. Consistent with these competing effects, both VV and VH correlations with Landsat LST are positive in 28 of 48 cities (plus signs in Fig. \ref{Var corrs}).

\subsection{Data Analysis: Autoencoder-Derived Error Types}

\begin{figure}[htbp!]
    \centering
    \includegraphics[width=1.0\linewidth]{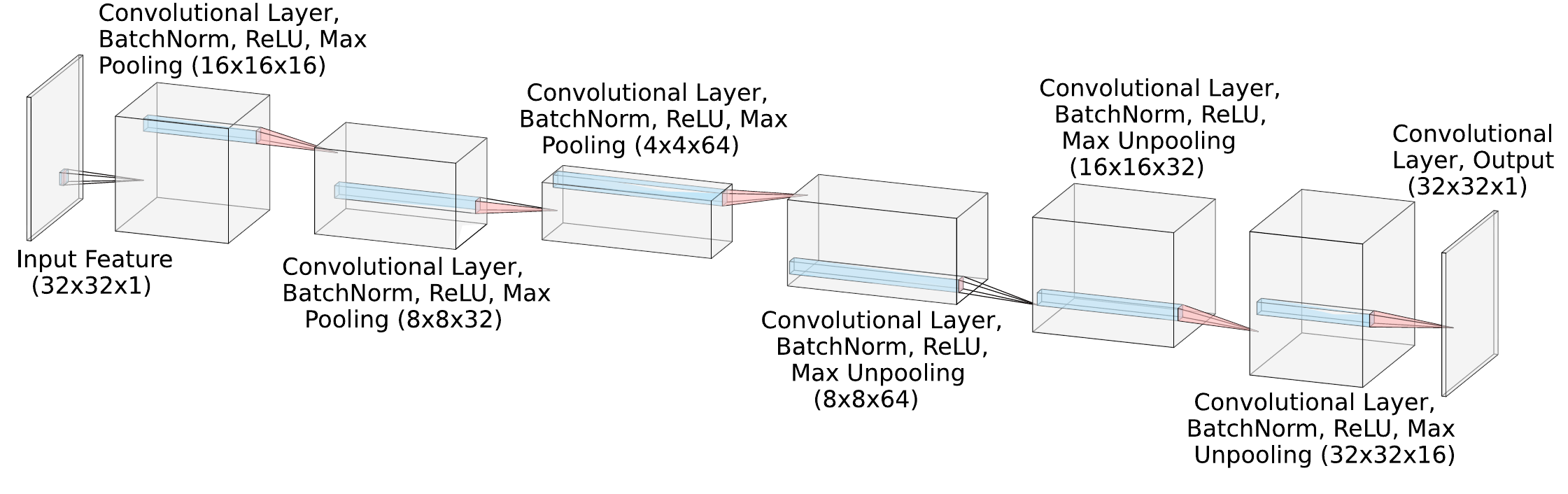}
    \caption{Schematic of the convolutional autoencoder architecture used to reconstruct Landsat LST images, made with \cite{lenail2020nn}.}
    \label{Autoencoder architecture}
\end{figure}

In the Landsat LST imagery, spatially structured artifacts and heterogeneous surface types (e.g., water and cloud-contaminated pixels) can introduce systematic patterns that may be inadvertently learned by machine learning algorithms. To identify which geographic features are poorly reconstructed from Landsat LST alone, we trained a city-specific convolutional autoencoder to compress and reconstruct Landsat LST images, using the resulting reconstruction errors to identify high-error surface types (Fig. \ref{Autoencoder architecture}). For each city, we partitioned the available Landsat LST images into 48\% training, 12\% validation, and 40\% test sets. Each $3000\times3000$ image was subdivided into $32\times32$ patches to reduce memory requirements and improve training efficiency.

\begin{figure}[htbp!]
    \centering
    \includegraphics[width=1.0\linewidth]{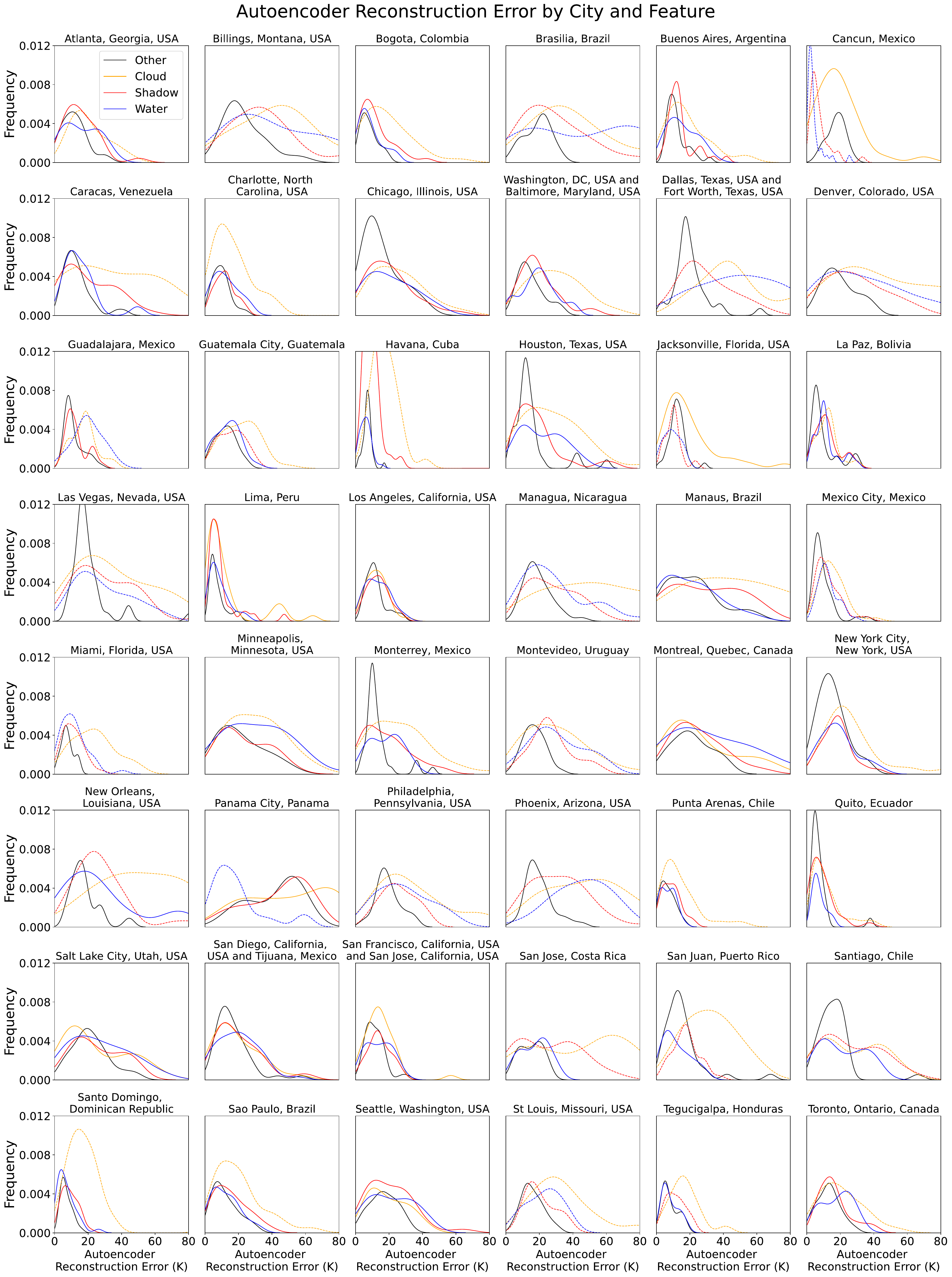}
    \caption{PDFs of autoencoder reconstruction error for Landsat LST, stratified by pixel category (cloud, cloud shadow, water, and other) as defined by the Landsat cloud mask, for each city in the dataset. PDF estimates are constructed from per-image mean reconstruction errors for each pixel category computed on a held-out test set. Dashed curves indicate distributions that differ significantly from the `other' category ($p < 0.05$) based on a Mann-Whitney U test (two-tailed).}
    \label{Feature reconstruction error}
\end{figure}

Autoencoders were trained for up to 20 epochs using the Adam optimizer with a learning rate of 0.001, and early stopping was triggered if the validation error increased for two consecutive epochs. The model architecture is shown in Fig. \ref{Autoencoder architecture} and detailed in Table \ref{CNN architecture table}. After training, reconstruction error was computed for each test image and summarized by pixel type using the Landsat cloud mask. Specifically, we grouped pixels into four types/categories and computed the mean reconstruction error for each category per image. The four categories are cloud, cloud shadow, water, and all remaining `other' pixels, which we interpret as predominantly clear-sky, non-water LST pixels. These per-image summary statistics were used to generate PDFs of reconstruction error by pixel category (Fig. \ref{Feature reconstruction error}).

To assess whether reconstruction error differs systematically across pixel categories, we performed two-sided Mann-Whitney U tests comparing the per-image mean reconstruction error distributions for each category (cloud, cloud shadow, and water) against the distribution for `other' pixels. Pixel categories with statistically significant differences ($p<0.05$) are indicated by dashed lines in Fig. \ref{Feature reconstruction error}. This analysis identifies which surface or atmospheric conditions are associated with larger reconstruction errors, providing evidence for feature-dependent error structure in reconstructed Landsat LST imagery across cities in the dataset. 

Across many cities, the 'other' pixel category exhibits lower reconstruction error, consistent with the pixel category typically comprising the largest fraction of each image, which provides the autoencoder with the most training examples for that category and yields more stable error estimates. By similar reasoning, coastal cities whose export grids contain substantial ocean area (e.g., Cancun, Havana, Jacksonville, and San Juan) would be expected to exhibit relatively low reconstruction error for 'water' pixels, since water constitutes a large, spatially coherent category in the respective scenes. This pattern largely holds, except for Los Angeles and New York City, where `water' pixels exhibit higher reconstruction errors, though these differences are not statistically significant. 

Panama City is notable in that `water' pixels exhibit substantially lower reconstruction error than any other pixel category. The higher errors for the other pixel categories may reflect greater spatial heterogeneity in Landsat LST within the city’s domain across seasons (Fig. \ref{LST std seasonal variability}), which could reduce reconstruction fidelity. Reconstruction error distributions also show similarities across regional climate contexts. For example, Las Vegas and Phoenix, both in deserts in the U.S. Southwest, exhibit a pronounced, unimodal error peak for `other' pixels, whereas other pixel category errors are comparatively flat. In contrast, San Diego, San Francisco, and Seattle, all with a Mediterranean climate on the U.S. West Coast, show broadly similar, moderately peaked distributions across pixel categories, with no pixel category exhibiting statistically significant differences.

\subsection{Data Limitations and Usage Notes}

Although LST and near-surface air temperature are often spatially correlated, their relationship varies with land cover, seasonality, and diurnal patterns \cite{naserikia2023land, good2017spatiotemporal, mildrexler2011global, sheng2017quantifying, jin2010land, sohrabinia2015spatio, cao2021within}. Users should therefore avoid interpreting LST as a direct proxy for near-surface air temperature, particularly in applications that require human-relevant exposure metrics. The thermal band used for the Landsat LST algorithm is Landsat TIRS band 10, which is observed at 100 m resolution. The 30 m pixel resolution of the L2 Landsat LST product is obtained through the auxiliary environmental input data to the processing algorithm \cite{LandsatCalVal}. Users should be aware of this resolution distinction in the Landsat LST product when performing fine-scale (i.e., neighborhood-level) analysis.

Cloud contamination further complicates analyses based on TIR observations. As discussed in Section \ref{mw_overview}, LWIR radiation is attenuated (through absorption and scattering) by clouds because its wavelength is not substantially larger than typical cloud and precipitation particle sizes. Consequently, LST estimates derived from TIR measurements may be biased low (cool) relative to the true surface temperature under cloudy or partially cloudy conditions. GOES BTs are also derived from LWIR measurements and are therefore also susceptible to cloud-related attenuation. While both products typically report cooler temperatures when clouds are present, they differ in their measurement definitions and processing. Specifically, GOES BTs represent TOA radiances, whereas Landsat LST is a surface retrieval that incorporates additional information (e.g., emissivity and a DEM) to account for surface properties and topographic effects, in addition to an atmospheric correction. The MW LST product included in Urban Heat MiniCubes may help mitigate cloud attenuation limitations, since microwave wavelengths are minimally attenuated by clouds; however, its $0.25\degree$ spatial resolution limits its utility for fine-scale urban analyses. We also note that the MW LST product provides a cloud-invariant temperature estimate derived from diurnal temperature-cycle behavior. Accordingly, these temperature variables should not be treated as interchangeable measurements. Extra care should be taken when performing analysis on cloudy images, as cloud edge detection from the FMask algorithm is known to be imperfect \cite{sui2022comparative}.

Nearest-neighbor resampling was applied during regridding operations when producing the dataset. This approach can introduce localized artifacts, particularly with lower-resolution instruments such as GOES ABI, when sharp spatial gradients can occur across land-water boundaries or heterogeneous urban surfaces. For example, in coastal cities during cooler months, water surfaces may be warmer than the adjacent land. Under these conditions, cartographic error and geolocation uncertainty (1.414 km; Section \ref{validation_section}) can lead to mixing between water and land pixels, resulting in anomalous temperature measurements. 

Sentinel-1 polarization availability is not uniform across acquisitions. Only a small subset of Sentinel-1 observations in the dataset contain HH and HV polarizations, since VV and VH are the standard dual-polarization measurements for IW mode over land. To maximize temporal alignment with Landsat acquisitions and preserve the closest-in-time SAR observation, we retain HH and HV images when available rather than restricting the dataset to VV and VH only. Users are discouraged from treating VV and HH, or VH and HV, as equivalent channels, as these polarizations generally represent distinct scattering responses and are not comparable on a pixel-by-pixel basis.

Additional datasets can supplement the Landsat surface reflectances in Urban Heat Minicubes. For example, the Harmonized Landsat and Sentinel-2 (HLS-2) provides harmonized surface reflectance data from OLI and Multi-Spectral Instrument (MSI) aboard Landsat and Sentinel-2 at 30 m resolution every 2-3 days \cite{ju2025harmonized}.

\section{Potential Use Cases}

Recent advances in AI have expanded the feasibility of working with large-scale Earth observation datasets, including high-volume remote sensing archives \cite{molina2023review, eyring2024pushing}. One prominent application is super-resolution, which aims to increase the spatial resolution of an input image by learning a mapping from coarse to fine scales \cite{ledig2017photo}. Remote sensing data are particularly well-suited for super-resolution problems due to the complementary sampling characteristics of polar-orbiting and geostationary satellite platforms. Geostationary sensors provide frequent observations at comparatively coarse spatial resolution, whereas polar-orbiting sensors provide higher-spatial-resolution observations at lower temporal frequency. This complementary structure enables learning-based downscaling, in which high-resolution polar-orbiting observations serve as reference targets for super-resolution of geostationary imagery, producing outputs with improved spatial detail while retaining high temporal coverage. Numerous studies have demonstrated the utility of super-resolution methods for remote sensing applications \cite{wang2022comprehensive}, including several focused on downscaling LST \cite{nguyen2022convolutional, lee2025guided}. However, further methodological improvements remain necessary to reliably recover urban microscale thermal variability, and datasets such as Urban Heat MiniCubes can help accelerate progress by providing paired observations across sensors and scales. Emerging Earth observation foundation models may also support super-resolution and related downscaling tasks by learning transferable latent representations from large, heterogeneous satellite data archives and multi-sensor data streams \cite{brown2025foundations, feng2025tessera}.

A second class of AI applications involves inpainting, which seeks to reconstruct missing or anomalous regions in imagery \cite{liu2018image}. In the remote sensing context, inpainting is particularly relevant for surface temperature products affected by cloud contamination, since clouds attenuate longwave infrared radiation and obscure surface conditions. When cloud-contaminated pixels are treated as missing, inpainting methods can infer plausible surface temperature values using information from surrounding clear sky pixels and broader spatial context. Prior work has applied inpainting strategies to sea surface temperature products \cite{dong2018inpainting, wei2023inpainting}, whereas comparatively fewer studies have focused on land surface temperature reconstruction \cite{huber2024deep}. Urban Heat MiniCubes provides a useful testbed for such methods by combining frequent geostationary observations with higher-resolution thermal imagery and microwave LST, enabling evaluation of gap-filling performance across heterogeneous urban landscapes.

Beyond technical applications, Urban Heat MiniCubes also support research on the societal and environmental dimensions of urban heat. Urban green space is associated with multiple public-health and environmental benefits, including increased physical activity, improved mental health outcomes, and cleaner local air quality \cite{wolch2014urban, elliott2020improving}. At the same time, communities most exposed to extreme heat frequently have fewer resources to implement heat mitigation interventions, including investment in green infrastructure \cite{li2024green, zhou2021urban}. Disparities in access to green space and heat-mitigation capacity have therefore been framed as an environmental justice issue. In many U.S. cities, these disparities may be linked to historical practices such as redlining, which contributed to long-term inequities in neighborhood-level investment and access to amenities \cite{chakraborty2023residential}. Given these human impacts, fine-scale characterization of intra-urban temperature variability is essential for understanding differential heat exposure, particularly in historically underserved communities.

\section{Data Availability} \label{DataAvailavility}

The full Urban Heat MiniCubes dataset is publicly available on the NSF National Center for Atmospheric Research (NCAR) Geoscience Data Exchange (GDEX) \cite{Urban_Heat_MiniCubes}.

All Landsat, Sentinel-1, and GOES imagery used in this dataset is open-source and available online. Landsat 8/9 imagery is available from the United States Geological Survey's website \cite{Landsat_dataset} in GeoTIFF format. Sentinel-1 imagery is available at the European Space Agency's Copernicus Data Space Ecosystem \cite{Sentinel1_dataset} in GeoTIFF format. GOES ABI image data can be accessed from the National Oceanic and Atmospheric Administration (NOAA) National Centers for Environmental Information's website \cite{GOES_CMI_dataset} in netCDF format.

The data sources for the MW LST product \cite{holmes2015diurnal} are available online: Microwave BT measurements for WindSat can be downloaded from NASA EarthData \cite{WindSat_BT}, and for AMSR2 from the National Snow and Ice Data Center \cite{AMSR2_BT}. BTs from older or decommissioned sensors, namely DMSP F13-16, TRMM TMI, and Aqua AMSR-E, are not readily available in their original form, but they have been provided as common calibrated BTs with the Global Precipitation Mission (GPM) by the NASA Goddard Earth Sciences Data and Information Services Center (GES DISC) \cite{F13_BT, F14_BT, F15_BT, F16_BT, TRMM_BT, AMSRE_BT}. Geostationary TIR LST observations from the Meteosat Second Generation satellite, used for construction of the diurnal temperature cycle, are provided by the European Organisation for the Exploitation of Meteorological Satellites (EUMETSAT) \cite{MSG_LST}. Reference reanalysis surface temperature is taken from NASA's Modern Era Reanalysis for Research and Applications (MERRA) \cite{MERRA_LST}.

\section{Code Availability}

All code used to create and process the Urban Heat MiniCubes dataset is publicly available \cite{project_Github_repo}.

\bibliography{sn-bibliography}

@STRING{AN        = "Astrophys.\ Norv."}

@STRING{GA        = "Geophysica"}

@STRING{JM        = "J.\ Meteor."}

@STRING{MA        = "Meteor.\ Appl."}

@STRING{MAP       = "Meteor.\ Atmos.\ Phys."}

@article{molina2023review,
  title={{A review of recent and emerging machine learning applications for climate variability and weather phenomena}},
  author={Molina, Maria J and O’Brien, Travis A and Anderson, Gemma and Ashfaq, Moetasim and Bennett, Katrina E and Collins, William D and Dagon, Katherine and Restrepo, Juan M and Ullrich, Paul A},
  journal={Artificial Intelligence for the Earth Systems},
  volume={2},
  number={4},
  pages={220086},
  year={2023},
  doi={10.1175/AIES-D-22-0086.1}
}

@article{kumar2024urban,
  title={{Urban heat mitigation by green and blue infrastructure: Drivers, effectiveness, and future needs}},
  author={Kumar, Prashant and Debele, Sisay E and Khalili, Soheila and Halios, Christos H and Sahani, Jeetendra and Aghamohammadi, Nasrin and de Fatima Andrade, Maria and Athanassiadou, Maria and Bhui, Kamaldeep and Calvillo, Nerea and others},
  journal={The Innovation},
  volume={5},
  number={2},
  year={2024},
  publisher={Elsevier},
doi={10.1016/j.xinn.2024.100588}
}

@incollection{murphy2006visible,
  title={The visible infrared imaging radiometer suite},
  author={Murphy, RE and Ardanuy, Phillip and Deluccia, Frank J and Clement, JE and Schueler, Carl F},
  booktitle={Earth Science Satellite Remote Sensing: Vol. 1: Science and Instruments},
  pages={199--223},
  year={2006},
  publisher={Springer},
  address={Berlin, Heidelberg},
  doi={10.1007/978-3-540-37293-6_11}
}

@article{eyring2024pushing,
  title={{Pushing the frontiers in climate modelling and analysis with machine learning}},
  author={Eyring, Veronika and Collins, William D and Gentine, Pierre and Barnes, Elizabeth A and Barreiro, Marcelo and Beucler, Tom and Bocquet, Marc and Bretherton, Christopher S and Christensen, Hannah M and Dagon, Katherine and others},
  journal={Nature Climate Change},
  volume={14},
  number={9},
  pages={916--928},
  year={2024},
  publisher={Nature Publishing Group UK London},
doi={10.1038/s41558-024-02095-y}
}

@article{brotzge2020technical,
  title={{A technical overview of the New York State Mesonet standard network}},
  author={Brotzge, Jerald A and Wang, J and Thorncroft, CD and Joseph, E and Bain, N and Bassill, N and Farruggio, N and Freedman, JM and Hemker Jr, K and Johnston, D and others},
  journal={Journal of Atmospheric and Oceanic Technology},
  volume={37},
  number={10},
  pages={1827--1845},
  year={2020},
doi={10.1175/JTECH-D-19-0220.1}
}

@article{ahmed2025AIHeat,
  title={{The urban heat Island effect: A review on predictive approaches using artificial intelligence models}},
  author={Ahmed, Ali Najah and AlDahoul, Nouar and Aziz, Nurhanani A and Huang, YF and Sherif, Mohsen and El-Shafie, Ahmed},
  journal={City and Environment Interactions},
  pages={100234},
  year={2025},
  publisher={Elsevier},
doi={10.1016/j.cacint.2025.100234}
}

@article{bai2021SARLULC,
  title={{Comprehensively analyzing optical and polarimetric SAR features for land-use/land-cover classification and urban vegetation extraction in highly-dense urban area}},
  author={Bai, Yunkun and Sun, Guangmin and Li, Yu and Ma, Peifeng and Li, Gang and Zhang, Yuanzhi},
  journal={International Journal of Applied Earth Observation and Geoinformation},
  volume={103},
  pages={102496},
  year={2021},
  publisher={Elsevier},
doi={10.1016/j.jag.2021.102496}
}

@article{beck2018present,
  title={{Present and future K{\"o}ppen-Geiger climate classification maps at 1-km resolution}},
  author={Beck, Hylke E and Zimmermann, Niklaus E and McVicar, Tim R and Vergopolan, Noemi and Berg, Alexis and Wood, Eric F},
  journal={Scientific data},
  volume={5},
  number={1},
  pages={1--12},
  year={2018},
  publisher={Nature Publishing Group},
doi={10.1038/sdata.2018.214}
}

@article{benali2012estimating,
  title={{Estimating air surface temperature in Portugal using MODIS LST data}},
  author={Benali, A and Carvalho, AC and Nunes, Jo{\~a}o Pedro and Carvalhais, Nuno and Santos, A},
  journal={Remote sensing of environment},
  volume={124},
  pages={108--121},
  year={2012},
  publisher={Elsevier},
doi={10.1016/j.rse.2012.04.024}
}

@inproceedings{brooks1973universal,
  title={{The universal transverse mercator grid}},
  author={Brooks, WD},
  booktitle={Proceedings of the Indiana Academy of Science},
  volume={83},
  pages={250--258},
  year={1973}
}

@article{brown2025foundations,
    author = {Brown, C and Kazmierski, M and Pasquarella V},
    title = {{AlphaEarth Foundations: An embedding field model for accurate and efficient global mapping from sparse label data}},
    journal = {},
    year = {2025},
doi={10.48550/arXiv.2507.22291}
}

@article{cao2021within,
  title={{Within-city spatial and temporal heterogeneity of air temperature and its relationship with land surface temperature}},
  author={Cao, Jie and Zhou, Weiqi and Zheng, Zhong and Ren, Tian and Wang, Weimin},
  journal={Landscape and Urban Planning},
  volume={206},
  pages={103979},
  year={2021},
  publisher={Elsevier},
doi={10.1016/j.landurbplan.2020.103979}
}

@article{chakraborty2023residential,
  title={{Residential segregation and outdoor urban moist heat stress disparities in the United States}},
  author={Chakraborty, TC and Newman, Andrew J and Qian, Yun and Hsu, Angel and Sheriff, Glenn},
  journal={One Earth},
  volume={6},
  number={6},
  pages={738--750},
  year={2023},
  publisher={Elsevier},
doi={10.1016/j.oneear.2023.05.016}
}

@article{chen2017tutorial,
  title={{A tutorial on kernel density estimation and recent advances}},
  author={Chen, Yen-Chi},
  journal={Biostatistics \& Epidemiology},
  volume={1},
  number={1},
  pages={161--187},
  year={2017},
  publisher={Taylor \& Francis},
doi={10.1080/24709360.2017.1396742}
}

@article{do2022comparison,
  title={{Comparison between air temperature and land surface temperature for the city of S{\~a}o Paulo, Brazil}},
  author={do Nascimento, Augusto Cezar Lima and Galvani, Emerson and Gobo, Jo{\~a}o Paulo Assis and Wollmann, C{\'a}ssio Arthur},
  journal={Atmosphere},
  volume={13},
  number={3},
  pages={491},
  year={2022},
  publisher={MDPI},
doi={10.3390/atmos13030491}
}

@article{dong2018inpainting,
  title={{Inpainting of remote sensing SST images with deep convolutional generative adversarial network}},
  author={Dong, Junyu and Yin, Ruiying and Sun, Xin and Li, Qiong and Yang, Yuting and Qin, Xukun},
  journal={IEEE geoscience and remote sensing letters},
  volume={16},
  number={2},
  pages={173--177},
  year={2018},
  publisher={IEEE},
doi={10.1109/LGRS.2018.2870880}
}

@article{elliott2020improving,
  title={{Improving City vitality through urban heat reduction with green infrastructure and design solutions: A systematic literature review}},
  author={Elliott, Helen and Eon, Christine and Breadsell, Jessica K},
  journal={Buildings},
  volume={10},
  number={12},
  pages={219},
  year={2020},
  publisher={MDPI},
doi={10.3390/buildings10120219}
}

@article{feng2025tessera,
  title={{Tessera: Temporal embeddings of surface spectra for earth representation and analysis}},
  author={Feng, Zhengpeng and Atzberger, Clement and Jaffer, Sadiq and Knezevic, Jovana and Sormunen, Silja and Young, Robin and Lisaius, Madeline C and Immitzer, Markus and Jackson, Toby and Ball, James and others},
  journal={arXiv preprint arXiv:2506.20380},
  year={2025},
  doi={10.48550/arXiv.2506.20380}
}

@article{foga2017cloud,
  title={{Cloud detection algorithm comparison and validation for operational Landsat data products}},
  author={Foga, Steve and Scaramuzza, Pat L and Guo, Song and Zhu, Zhe and Dilley Jr, Ronald D and Beckmann, Tim and Schmidt, Gail L and Dwyer, John L and Hughes, M Joseph and Laue, Brady},
  journal={Remote sensing of environment},
  volume={194},
  pages={379--390},
  year={2017},
  publisher={Elsevier},
doi={10.1016/j.rse.2017.03.026}
}

@article{gavsparovic2020comparative,
  title={{Comparative assessment of machine learning methods for urban vegetation mapping using multitemporal sentinel-1 imagery}},
  author={Ga{\v{s}}parovi{\'c}, Mateo and Dobrini{\'c}, Dino},
  journal={Remote Sensing},
  volume={12},
  number={12},
  pages={1952},
  year={2020},
  publisher={MDPI},
doi={10.3390/rs12121952}
}

@article{good2017spatiotemporal,
  title={{A spatiotemporal analysis of the relationship between near-surface air temperature and satellite land surface temperatures using 17 years of data from the ATSR series}},
  author={Good, Elizabeth J and Ghent, Darren J and Bulgin, Claire E and Remedios, John J},
  journal={Journal of Geophysical Research: Atmospheres},
  volume={122},
  number={17},
  pages={9185--9210},
  year={2017},
  publisher={Wiley Online Library},
doi={10.1002/2017JD026880}
}

@article{gorelick2017google,
  title={{Google Earth Engine: Planetary-scale geospatial analysis for everyone}},
  author={Gorelick, Noel and Hancher, Matt and Dixon, Mike and Ilyushchenko, Simon and Thau, David and Moore, Rebecca},
  journal={Remote sensing of Environment},
  volume={202},
  pages={18--27},
  year={2017},
  publisher={Elsevier},
doi={10.1016/j.rse.2017.06.031}
}

@article{harries2008far,
  title={{The far-infrared Earth}},
  author={Harries, John and Carli, Bruno and Rizzi, Rolando and Serio, Carmine and Mlynczak, M and Palchetti, Luca and Maestri, Tiziano and Brindley, H and Masiello, Guido},
  journal={Reviews of Geophysics},
  volume={46},
  number={4},
  year={2008},
  publisher={Wiley Online Library},
doi={10.1029/2007RG000233}
}

@article{hart2009quantifying,
  title={{Quantifying the influence of land-use and surface characteristics on spatial variability in the urban heat island}},
  author={Hart, Melissa A and Sailor, David J},
  journal={Theoretical and applied climatology},
  volume={95},
  number={3},
  pages={397--406},
  year={2009},
  publisher={Springer},
doi={10.1007/s00704-008-0017-5}
}

@article{holmes2015diurnal,
  title={{Diurnal temperature cycle as observed by thermal infrared and microwave radiometers}},
  author={Holmes, TRH and Crow, WT and Hain, C and Anderson, MC and Kustas, WP},
  journal={Remote Sensing of Environment},
  volume={158},
  pages={110--125},
  year={2015},
  publisher={Elsevier},
doi={10.1016/j.rse.2014.10.031}
}

@article{holmes2018microwave,
  title={{Microwave implementation of two-source energy balance approach for estimating evapotranspiration}},
  author={Holmes, Thomas RH and Hain, Christopher R and Crow, Wade T and Anderson, Martha C and Kustas, William P},
  journal={Hydrology and earth system sciences},
  volume={22},
  number={2},
  pages={1351--1369},
  year={2018},
  publisher={Copernicus GmbH},
doi={10.5194/hess-22-1351-2018}
}

@article{huber2024deep,
  title={{Deep interpolation of remote sensing land surface temperature data with partial convolutions}},
  author={Huber, Florian and Schulz, Stefan and Steinhage, Volker},
  journal={Sensors},
  volume={24},
  number={5},
  pages={1604},
  year={2024},
  publisher={MDPI},
doi={10.3390/s24051604}
}

@article{imran2021impact,
  title={{Impact of land cover changes on land surface temperature and human thermal comfort in Dhaka city of Bangladesh}},
  author={Imran, HM and Hossain, Anwar and Islam, AKM Saiful and Rahman, Ataur and Bhuiyan, Md Abul Ehsan and Paul, Supria and Alam, Akramul},
  journal={Earth Systems and Environment},
  volume={5},
  number={3},
  pages={667--693},
  year={2021},
  publisher={Springer},
doi={10.1007/s41748-021-00243-4}
}

@article{jin2010land,
  title={{Land surface skin temperature climatology: benefitting from the strengths of satellite observations}},
  author={Jin, Menglin and Dickinson, Robert E},
  journal={Environmental research letters},
  volume={5},
  number={4},
  pages={044004},
  year={2010},
  publisher={IOP Publishing},
doi={10.1088/1748-9326/5/4/044004}
}

@article{jones2007satellite,
  title={{Satellite microwave remote sensing of boreal and arctic soil temperatures from AMSR-E}},
  author={Jones, Lucas A and Kimball, John S and McDonald, Kyle C and Chan, Steven Tsz K and Njoku, Eni G and Oechel, Walter C},
  journal={IEEE Transactions on Geoscience and Remote Sensing},
  volume={45},
  number={7},
  pages={2004--2018},
  year={2007},
  publisher={IEEE},
doi={10.1109/TGRS.2007.898436}
}

@article{ju2025harmonized,
  title={{The Harmonized Landsat and Sentinel-2 version 2.0 surface reflectance dataset}},
  author={Ju, Junchang and Zhou, Qiang and Freitag, Brian and Roy, David P and Zhang, Hankui K and Sridhar, Madhu and Mandel, John and Arab, Saeed and Schmidt, Gail and Crawford, Christopher J and others},
  journal={Remote Sensing of Environment},
  volume={324},
  pages={114723},
  year={2025},
  publisher={Elsevier},
  doi={10.1016/j.rse.2025.114723}
}

@article{koukiou2024sar,
  title={{SAR Features and Techniques for Urban Planning—A Review}},
  author={Koukiou, Georgia},
  journal={Remote Sensing},
  volume={16},
  number={11},
  pages={1923},
  year={2024},
  publisher={MDPI},
doi={10.3390/rs16111923}
}

@inproceedings{ledig2017photo,
  title={{Photo-realistic single image super-resolution using a generative adversarial network}},
  author={Ledig, Christian and Theis, Lucas and Husz{\'a}r, Ferenc and Caballero, Jose and Cunningham, Andrew and Acosta, Alejandro and Aitken, Andrew and Tejani, Alykhan and Totz, Johannes and Wang, Zehan and others},
  booktitle={Proceedings of the IEEE conference on computer vision and pattern recognition},
  pages={4681--4690},
  year={2017}
}

@article{lee2025guided,
  title={{Guided Super Resolution of Land Surface Temperature Using Multi-Satellite Imageries}},
  author={Lee, Sunju and Choi, Yeji and Choi, Beomkyu and Seo, Junghoon and Song, Minki and Sohn, Eunha and Ahn, Sewong},
  journal={IEEE Transactions on Geoscience and Remote Sensing},
  year={2025},
  publisher={IEEE},
doi={10.1109/TGRS.2025.3572460}
}

@article{li2024green,
  title={{Green spaces provide substantial but unequal urban cooling globally}},
  author={Li, Yuxiang and Svenning, Jens-Christian and Zhou, Weiqi and Zhu, Kai and Abrams, Jesse F and Lenton, Timothy M and Ripple, William J and Yu, Zhaowu and Teng, Shuqing N and Dunn, Robert R and others},
  journal={Nature communications},
  volume={15},
  number={1},
  pages={7108},
  year={2024},
  publisher={Nature Publishing Group UK London},
doi={10.1038/s41467-024-51355-0}
}

@article{lindsey2024geoxo,
  title={{GeoXO: NOAA’s future geostationary satellite system}},
  author={Lindsey, Daniel T and Heidinger, Andrew K and Sullivan, Pamela C and McCorkel, Joel and Schmit, Timothy J and Tomlinson, Michelle and Vandermeulen, Ryan and Frost, Gregory J and Kondragunta, Shobha and Rudlosky, Scott},
  journal={Bulletin of the American Meteorological Society},
  volume={105},
  number={3},
  pages={E660--E679},
  year={2024},
  publisher={American Meteorological Society},
doi={10.1175/BAMS-D-23-0048.1}
}

@inproceedings{liu2018image,
  title={{Image inpainting for irregular holes using partial convolutions}},
  author={Liu, Guilin and Reda, Fitsum A and Shih, Kevin J and Wang, Ting-Chun and Tao, Andrew and Catanzaro, Bryan},
  booktitle={Proceedings of the European conference on computer vision (ECCV)},
  pages={85--100},
  year={2018}
}

@article{malakar2018LST,
  title={{An operational land surface temperature product for Landsat thermal data: Methodology and validation}},
  author={Malakar, Nabin K and Hulley, Glynn C and Hook, Simon J and Laraby, Kelly and Cook, Monica and Schott, John R},
  journal={IEEE Transactions on Geoscience and Remote Sensing},
  volume={56},
  number={10},
  pages={5717--5735},
  year={2018},
  publisher={IEEE},
doi={10.1109/TGRS.2018.2824828}
}

@article{mcknight2010mann,
  title={{Mann-whitney U test}},
  author={McKnight, Patrick E and Najab, Julius},
  journal={The Corsini encyclopedia of psychology},
  pages={1--1},
  year={2010},
  publisher={Wiley Online Library},
doi={10.1002/9780470479216.corpsy0524}
}

@article{mcpherson2007statewide,
  title={{Statewide monitoring of the mesoscale environment: A technical update on the Oklahoma Mesonet}},
  author={McPherson, Renee A and Fiebrich, Christopher A and Crawford, Kenneth C and Kilby, James R and Grimsley, David L and Martinez, Janet E and Basara, Jeffrey B and Illston, Bradley G and Morris, Dale A and Kloesel, Kevin A and others},
  journal={Journal of Atmospheric and Oceanic Technology},
  volume={24},
  number={3},
  pages={301--321},
  year={2007},
doi={10.1175/JTECH1976.1}
}

@article{memon2009investigation,
  title={{An investigation of urban heat island intensity (UHII) as an indicator of urban heating}},
  author={Memon, Rizwan Ahmed and Leung, Dennis YC and Liu, Chun-Ho},
  journal={Atmospheric Research},
  volume={94},
  number={3},
  pages={491--500},
  year={2009},
  publisher={Elsevier},
doi={10.1016/j.atmosres.2009.07.006}
}

@article{mildrexler2011global,
  title={{A global comparison between station air temperatures and MODIS land surface temperatures reveals the cooling role of forests}},
  author={Mildrexler, David J and Zhao, Maosheng and Running, Steven W},
  journal={Journal of Geophysical Research: Biogeosciences},
  volume={116},
  number={G3},
  year={2011},
  publisher={Wiley Online Library},
doi={10.1029/2010JG001486}
}

@article{moore1997transverse,
  title={{Transverse mercator projections and us geological survey digital products}},
  author={Moore, Laurence},
  journal={US Geological Survey, Professional Paper},
  year={1997}
}

@article{mitchell2015landscapes,
  title={{Landscapes of thermal inequity: disproportionate exposure to urban heat in the three largest US cities}},
  author={Mitchell, Bruce C and Chakraborty, Jayajit},
  journal={Environmental Research Letters},
  volume={10},
  number={11},
  pages={115005},
  year={2015},
  publisher={IOP Publishing},
doi={10.1088/1748-9326/10/11/115005}
}

@article{mohamed2024AIHeat,
  title={{Artificial intelligence for predicting urban heat island effect and optimising land use/land cover for mitigation: Prospects and recent advancements}},
  author={Mohamed, Omar YA and Zahidi, Izni},
  journal={Urban Climate},
  volume={55},
  pages={101976},
  year={2024},
  publisher={Elsevier},
doi={10.1016/j.uclim.2024.101976}
}

@article{naserikia2023land,
  title={{Land surface and air temperature dynamics: The role of urban form and seasonality}},
  author={Naserikia, Marzie and Hart, Melissa A and Nazarian, Negin and Bechtel, Benjamin and Lipson, Mathew and Nice, Kerry A},
  journal={Science of The Total Environment},
  volume={905},
  pages={167306},
  year={2023},
  publisher={Elsevier},
doi={10.1016/j.scitotenv.2023.167306}
}

@inproceedings{nguyen2022convolutional,
  title={{Convolutional neural network modelling for modis land surface temperature super-resolution}},
  author={Nguyen, Binh Minh and Tian, Ganglin and Vo, Minh-Triet and Michel, Aur{\'e}lie and Corpetti, Thomas and Granero-Belinchon, Carlos},
  booktitle={2022 30th European Signal Processing Conference (EUSIPCO)},
  pages={1806--1810},
  year={2022},
  organization={IEEE},
 doi={10.23919/EUSIPCO55093.2022.9909569}
}

@article{ni2024review,
  title={{A Review of Research on Cloud Detection Methods for Hyperspectral Infrared Radiances}},
  author={Ni, Zhuoya and Wu, Mengdie and Lu, Qifeng and Huo, Hongyuan and Wu, Chunqiang and Liu, Ruixia and Wang, Fu and Xu, Xiaoying},
  journal={Remote Sensing},
  volume={16},
  number={24},
  pages={4629},
  year={2024},
  publisher={MDPI},
doi={10.3390/rs16244629}
}

@article{palchetti2020FIR,
  title={{unique far-infrared satellite observations to better understand how Earth radiates energy to space}},
  author={Palchetti, L and Brindley, H and Bantges, R and Buehler, SA and Camy-Peyret, C and Carli, B and Cortesi, U and Del Bianco, S and Di Natale, G and Dinelli, BM and others},
  journal={Bulletin of the American meteorological society},
  volume={101},
  number={12},
  pages={E2030--E2046},
  year={2020},
doi={10.1175/BAMS-D-19-0322.1}
}

@article{parastatidis2017LST,
  title={{Online global land surface temperature estimation from Landsat}},
  author={Parastatidis, David and Mitraka, Zina and Chrysoulakis, Nektrarios and Abrams, Michael},
  journal={Remote sensing},
  volume={9},
  number={12},
  pages={1208},
  year={2017},
  publisher={MDPI},
doi={10.3390/rs9121208}
}

@article{peel2007updated,
  title={{Updated world map of the K{\"o}ppen-Geiger climate classification}},
  author={Peel, Murray C and Finlayson, Brian L and McMahon, Thomas A},
  journal={Hydrology and earth system sciences},
  volume={11},
  number={5},
  pages={1633--1644},
  year={2007},
  publisher={Copernicus GmbH},
doi={10.5194/hess-11-1633-2007}
}

@article{prigent2016toward,
  title={{Toward “all weather,” long record, and real-time land surface temperature retrievals from microwave satellite observations}},
  author={Prigent, C and Jimenez, C and Aires, Filipe},
  journal={Journal of Geophysical Research: Atmospheres},
  volume={121},
  number={10},
  pages={5699--5717},
  year={2016},
  publisher={Wiley Online Library},
doi={10.1002/2015JD024402}
}

@article{qi2022HeatDecision,
  title={{A decision-making framework to support urban heat mitigation by local governments}},
  author={Qi, Jinda and Ding, Lan and Lim, Samsung},
  journal={Resources, Conservation and Recycling},
  volume={184},
  pages={106420},
  year={2022},
  publisher={Elsevier},
doi={10.1016/j.resconrec.2022.106420}
}

@article{rizwan2008review,
  title={{A review on the generation, determination and mitigation of Urban Heat Island}},
  author={Rizwan, Ahmed Memon and Dennis, Leung YC and others},
  journal={Journal of environmental sciences},
  volume={20},
  number={1},
  pages={120--128},
  year={2008},
  publisher={Elsevier},
doi={10.1016/S1001-0742(08)60019-4}
}

@article{sauer2011mapping,
  title={{Three-dimensional imaging and scattering mechanism estimation over urban scenes using dual-baseline polarimetric InSAR observations at L-band}},
  author={Sauer, Stefan and Ferro-Famil, Laurent and Reigber, Andreas and Pottier, Eric},
  journal={IEEE Transactions on Geoscience and Remote Sensing},
  volume={49},
  number={11},
  pages={4616--4629},
  year={2011},
  publisher={IEEE},
doi={10.1109/TGRS.2011.2147321}
}

@article{scheffler2020harmonization,
  title={{Spectral harmonization and red edge prediction of Landsat-8 to Sentinel-2 using land cover optimized multivariate regressors}},
  author={Scheffler, Daniel and Frantz, David and Segl, Karl},
  journal={Remote Sensing of Environment},
  volume={241},
  pages={111723},
  year={2020},
  publisher={Elsevier},
doi={10.1016/j.rse.2020.111723}
}

@article{schmetz2002introduction,
  title={{An introduction to Meteosat second generation (MSG)}},
  author={Schmetz, Johannes and Pili, Paolo and Tjemkes, Stephen and Just, Dieter and Kerkmann, Jochen and Rota, Sergio and Ratier, Alain},
  journal={Bulletin of the American Meteorological Society},
  volume={83},
  number={7},
  pages={977--992},
  year={2002},
  publisher={American Meteorological Society},
doi={10.1175/1520-0477(2002)083<0977:AITMSG>2.3.CO;2}
}

@article{schmit2017closer,
  title={{A closer look at the ABI on the GOES-R series}},
  author={Schmit, Timothy J and Griffith, Paul and Gunshor, Mathew M and Daniels, Jaime M and Goodman, Steven J and Lebair, William J},
  journal={Bulletin of the American Meteorological Society},
  volume={98},
  number={4},
  pages={681--698},
  year={2017},
doi={10.1175/BAMS-D-15-00230.1}
}

@article{schuler2002surface,
  title={{Surface roughness and slope measurements using polarimetric SAR data}},
  author={Schuler, Dale L and Lee, Jong-Sen and Kasilingam, Dayalan and Nesti, Giuseppe},
  journal={IEEE Transactions on Geoscience and Remote Sensing},
  volume={40},
  number={3},
  pages={687--698},
  year={2002},
  publisher={IEEE},
doi={10.1109/TGRS.2002.1000328}
}

@article{scott1979optimal,
  title={{On optimal and data-based histograms}},
  author={Scott, David W},
  journal={Biometrika},
  volume={66},
  number={3},
  pages={605--610},
  year={1979},
  publisher={Oxford University Press},
doi={10.1093/biomet/66.3.605}
}

@article{sheng2017quantifying,
  title={{Quantifying the spatial and temporal relationship between air and land surface temperatures of different land-cover types in Southeastern China}},
  author={Sheng, Yanling and Liu, Xiaoping and Yang, Xuchao and Xin, Qinchuan and Deng, Chengbin and Li, Xia},
  journal={International Journal of Remote Sensing},
  volume={38},
  number={4},
  pages={1114--1136},
  year={2017},
  publisher={Taylor \& Francis},
doi={10.1080/01431161.2017.1280629}
}

@article{smirnov2019infrared,
  title={{Infrared Radiation in the Energetics of the Atmosphere}},
  author={Smirnov, BM},
  journal={High temperature},
  volume={57},
  number={4},
  pages={573--595},
  year={2019},
  publisher={Springer},
doi={10.1134/S0018151X19040199}
}

@article{sohrabinia2015spatio,
  title={{Spatio-temporal analysis of the relationship between LST from MODIS and air temperature in New Zealand}},
  author={Sohrabinia, M and Zawar-Reza, P and Rack, W},
  journal={Theoretical and applied climatology},
  volume={119},
  number={3},
  pages={567--583},
  year={2015},
  publisher={Springer},
doi={10.1007/s00704-014-1106-2}
}

@misc{lenail2020nn,
  title={NN-SVG: Publication-ready nn-architecture schematics},
  author={Lenail, Alex},
  year={2020},
  publisher={URl: http://alexlenail. me/NN-SVG}
}

@article{sui2022comparative,
  title={{Comparative analysis of several typical Landsat 8 OLI cloud detection methods}},
  author={Sui, Songman and Sun, Lin},
  journal={Remote Sensing},
  volume={14},
  number={3},
  pages={719},
  year={2022},
  publisher={MDPI},
  doi={10.3390/rs14030719}
}

@article{torres2012gmes,
  title={{GMES Sentinel-1 mission}},
  author={Torres, Ramon and Snoeij, Paul and Geudtner, Dirk and Bibby, David and Davidson, Malcolm and Attema, Evert and Potin, Pierre and Rommen, Bj{\"O}rn and Floury, Nicolas and Brown, Mike and others},
  journal={Remote sensing of environment},
  volume={120},
  pages={9--24},
  year={2012},
  publisher={Elsevier},
doi={10.1016/j.rse.2011.05.028}
}

@article{verhulst2025AIReady,
  title={{Moving Toward the FAIR-R principles: Advancing AI-Ready Data}},
  author={Verhulst, Stefaan and Zahuranec, Andrew J and Chafetz, Hannah},
  journal={Available at SSRN},
  year={2025},
doi={10.2139/ssrn.5164337}
}

@article{wang2019comparison,
  title={{Comparison of three algorithms for the retrieval of land surface temperature from Landsat 8 images}},
  author={Wang, Lei and Lu, Yao and Yao, Yunlong},
  journal={Sensors},
  volume={19},
  number={22},
  pages={5049},
  year={2019},
  publisher={MDPI},
doi={10.3390/s19225049}
}

@article{wang2022comprehensive,
  title={{A comprehensive review on deep learning based remote sensing image super-resolution methods}},
  author={Wang, Peijuan and Bayram, Bulent and Sertel, Elif},
  journal={Earth-Science Reviews},
  volume={232},
  pages={104110},
  year={2022},
  publisher={Elsevier},
doi={10.1016/j.earscirev.2022.104110}
}

@article{wang2025AIReady,
  title={{What Makes Data AI-Ready?}},
  author={Wang, Jingyang and Legner, Christine},
  journal={ECIS 2025 TREOs},
  year={2025},
  url={https://aisel.aisnet.org/treos_ecis2025/39}
}

@inproceedings{wkeglarczyk2018kernel,
  title={{Kernel density estimation and its application}},
  author={Weglarczyk, Stanislaw},
  booktitle={ITM web of conferences},
  volume={23},
  pages={00037},
  year={2018},
  organization={EDP Sciences},
doi={10.1051/itmconf/20182300037}
}

@inproceedings{wei2023inpainting,
  title={{Inpainting of remote sensing sea surface temperature image with multi-scale physical constraints}},
  author={Wei, Qichen and Zuo, Zijie and Nie, Jie and Du, Jiahao and Diao, Yaning and Ye, Min and Liang, Xinyue},
  booktitle={2023 IEEE International Conference on Multimedia and Expo (ICME)},
  pages={492--497},
  year={2023},
  organization={IEEE},
 doi={10.1109/ICME55011.2023.00091}
}

@article{wilkinson2016fair,
  title={{The FAIR Guiding Principles for scientific data management and stewardship}},
  author={Wilkinson, Mark D and Dumontier, Michel and Aalbersberg, IJsbrand Jan and Appleton, Gabrielle and Axton, Myles and Baak, Arie and Blomberg, Niklas and Boiten, Jan-Willem and da Silva Santos, Luiz Bonino and Bourne, Philip E and others},
  journal={Scientific data},
  volume={3},
  number={1},
  pages={1--9},
  year={2016},
  publisher={Nature Publishing Group},
doi={10.1038/sdata.2016.18}
}

@article{wolch2014urban,
  title={{Urban green space, public health, and environmental justice: The challenge of making cities ‘just green enough’}},
  author={Wolch, Jennifer R and Byrne, Jason and Newell, Joshua P},
  journal={Landscape and urban planning},
  volume={125},
  pages={234--244},
  year={2014},
  publisher={Elsevier},
doi={10.1016/j.landurbplan.2014.01.017}
}

@article{wulder2022fifty,
  title={{Fifty years of Landsat science and impacts}},
  author={Wulder, Michael A and Roy, David P and Radeloff, Volker C and Loveland, Thomas R and Anderson, Martha C and Johnson, David M and Healey, Sean and Zhu, Zhe and Scambos, Theodore A and Pahlevan, Nima and others},
  journal={Remote Sensing of Environment},
  volume={280},
  pages={113195},
  year={2022},
  publisher={Elsevier},
doi={10.1016/j.rse.2022.113195}
}

@article{zhang2016mapping,
  title={{Mapping urban impervious surface with dual-polarimetric SAR data: An improved method}},
  author={Zhang, Hongsheng and Lin, Hui and Li, Yu and Zhang, Yuanzhi and Fang, Chaoyang},
  journal={Landscape and urban planning},
  volume={151},
  pages={55--63},
  year={2016},
  publisher={Elsevier},
doi={10.1016/j.landurbplan.2016.03.009}
}

@article{young2017survival,
  title={{A survival guide to Landsat preprocessing}},
  author={Young, Nicholas E and Anderson, Ryan S and Chignell, Stephen M and Vorster, Anthony G and Lawrence, Rick and Evangelista, Paul H},
  journal={Ecology},
  volume={98},
  number={4},
  pages={920--932},
  year={2017},
  publisher={Wiley Online Library},
doi={10.1002/ecy.1730}
}

@article{zhao2020assessing,
  title={{Assessing the thermal contributions of urban land cover types}},
  author={Zhao, Jiacheng and Zhao, Xiang and Liang, Shunlin and Zhou, Tao and Du, Xiaozheng and Xu, Peipei and Wu, Donghai},
  journal={Landscape and Urban Planning},
  volume={204},
  pages={103927},
  year={2020},
  publisher={Elsevier},
doi={10.1016/j.landurbplan.2020.103927}
}

@article{zhou2021urban,
  title={{Urban tree canopy has greater cooling effects in socially vulnerable communities in the US}},
  author={Zhou, Weiqi and Huang, Ganlin and Pickett, Steward TA and Wang, Jing and Cadenasso, Mary L and McPhearson, Timon and Grove, J Morgan and Wang, Jia},
  journal={One Earth},
  volume={4},
  number={12},
  pages={1764--1775},
  year={2021},
  publisher={Elsevier},
doi={10.1016/j.oneear.2021.11.010 }
}

@misc{esipaiready,
author={ESIP},
title={{Are Your Data Ready? Take Stock with ESIP’s New AI-Ready Checklist}},
year={2022},
url={https://www.esipfed.org/checklist-ai-ready-data/},
note = {Accessed on 6 August 2025}
}

@techreport{GOEShandbook,
author={Valenti, Jim},
title={{GOES-R SERIES PRODUCT DEFINITION AND USERS' GUIDE}},
institution={United States Department of Commerce and National Oceanic and Atmospheric Administration and NOAA Satellite and Information Service National Aeronautics and Space Administration},
year={2019}
}

@techreport{LandsatHandbook,
author={Ihlen, Vaughn},
title={{Landsat 8 (L8) Data Users Handbook}},
institution={United States Geological Survey},
year={2019}
}

@techreport{LandsatCalVal,
author={Anderson, Cody},
title={{Landsat 8-9 Calibration and Validation (Cal/Val) Algorithm Description Document (ADD)}},
institution={United States Geological Survey},
year={2025}
}

@techreport{GOESperformance,
author={Wu, Xiangqian and Shmit, Tim},
title={{Product Performance Guide for Data Users of GOES-17 ABI Level 1b and Cloud and Moisture Imagery (CMI) Products}},
institution={National Oceanic and Atmospheric Administration},
year={2020} 
}

@techreport{GOESLSTalgorithm,
    author = {Sun, Donglian and Fang, Li and Yu, Yunyue},
    title = {{GOES LST ALGORITHM THEORETICAL BASIS DOCUMENT VERSION 3.0}},
    institution = {NOAA NESDIS STAR},
    year = {2012}
}

@misc{USCensus,
author={U.S. Census Bureau},
title={{2020 Census Urban Areas Facts}},
year={2023},
url={https://www.census.gov/programs-surveys/geography/guidance/geo-areas/urban-rural/2020-ua-facts.html},
note={Accessed on 6 August 2025}
}

@misc{Sentiwiki,
author={ESA},
title={{Sentinel-1. SentiWiki Home.}},
year={2025},
url={https://sentiwiki.copernicus.eu/web/sentinel-1},
note={Accessed on 6 August 2025}
}

@book{petty2006first,
  title={A first course in atmospheric radiation},
  edition={2},
  author={Petty, Grant W},
  year={2006},
  publisher={Sundog Publishing},
chapter={12},
address={Madison, WI}
}

@misc{GOES_CMI_dataset,
author={{GOES-R Algorithm Working Group; NOAA Geostationary Operation Environmental Satellite-R Series}},
title={{NOAA Geostationary Operation Environmental Satellite-R Series (GOES-R Series) Advanced Baseline Imager (ABI) Level 2 Cloud and Moisture Imagery Products (CMIP)}},
year={2017},
doi={10.7289/V5736P36}
}

@misc{Sentinel1_dataset,
author={{European Space Agency}},
title={{Sentinel-1 Ground Range Detected Product}},
year={2014},
url={https://browser.dataspace.copernicus.eu/}
}

@misc{Landsat_dataset,
author={{U.S. Geological Survey's Earth Resources Observation and Science (EROS) Center}},
title={{Landsat 8-9 Operational Land Imager / Thermal Infrared Sensor Level-2, Collection 2, Tier 1}},
year={2020},
doi={10.5066/P9OGBGM6}
}

@misc{project_Github_repo,
author={Starfeldt, Jonathan and Molina, Maria J and Kerr, Alexander and Yang, Adam},
title={{uhminicubes}},
year={2026},
doi={10.5281/zenodo.20600581}
}

@misc{CF_conventions,
author={Eaton, Brian and Gregory, Jonathan and Drach, Bob and Taylor, Karl and Hankin, Steve and others},
title={{NetCDF Climate and Forecast (CF) Metadata Conventions (1.13)}},
year={2025},
doi={10.5281/zenodo.17801666}
}

@misc{MSG_LST,
author={{LSA SAF}},
title={{Land Surface Temperature Data Record - MSG}},
year={2019},
doi={10.15770/EUM_SAF_LSA_0001}
}

@misc{F13_BT,
author={Wesley Berg},
title={{GPM SSMI on F13 Common Calibrated Brightness Temperatures L1C 1.5 hours 13 km V07}},
year={2021},
doi={10.5067/GPM/SSMI/F13/1C/07}
}

@misc{F14_BT,
author={Wesley Berg},
title={{GPM SSMI on F14 Common Calibrated Brightness Temperatures L1C 1.5 hours 13 km V07}},
year={2021},
doi={10.5067/GPM/SSMI/F14/1C/07}
}

@misc{F15_BT,
author={Wesley Berg},
title={{GPM SSMI on F15 Common Calibrated Brightness Temperatures L1C 1.5 hours 13 km V07}},
year={2021},
doi={10.5067/GPM/SSMI/F15/1C/07}
}

@misc{F16_BT,
author={Wesley Berg},
title={{GPM SSMIS on F16 Common Calibrated Brightness Temperatures L1C 1.5 hours 12 km V08}},
year={2026},
doi={10.5067/GPM/SSMIS/F16/1C/08}
}

@misc{TRMM_BT,
author={{GPM Science Team}},
title={{GPM TMI on TRMM Common Calibrated Brightness Temperatures L1C 1.5 hours 13 km V07}},
year={2021},
doi={10.5067/GPM/TMI/TRMM/1C/07}
}

@misc{AMSRE_BT,
author={Wesley Berg},
title={{GPM AMSR-E on AQUA Common Calibrated Brightness Temperatures L1C 1.5 hours 10.5 km V07}},
year={2021},
doi={10.5067/GPM/AMSRE/AQUA/1/07}
}

@misc{AMSR2_BT,
author={Meier, W. N. and Markus, T and Comiso, J. C.},
title={{AMSR-E/AMSR2 Unified L3 Daily 12.5 km Brightness Temperatures, Sea Ice Concentration, Motion \& Snow Depth Polar Grids. (AU\_SI12, Version 1)}},
year={2018},
doi={10.5067/RA1MIJOYPK3P}
}

@misc{WindSat_BT,
author={{Remote Sensing Systems}},
title={{RSS WindSat L1C Calibrated TB Version 8}},
year={2022},
doi={10.5067/WSA80-1CRTB}
}

@misc{MERRA_LST,
author={{Global Modeling and Assimilation Office (GMAO)}},
title={{MERRA-2 inst1\_2d\_asm\_Nx: 2d,1-Hourly,Instantaneous,Single-Level,Assimilation,Single-Level Diagnostics V5.12.4}},
year={2015},
doi={10.5067/3Z173KIE2TPD}
}

@misc{Urban_Heat_MiniCubes,
author={Starfeldt, Jonathan and Molina, Maria J and Kerr, Alexander and Yang, Adam and Holmes, Thomas RH and Hain, Christopher R},
title={{Urban Heat MiniCubes}},
year={2026},
doi={10.5065/GNGZ-2510}
}

\section*{Acknowledgments}
Processing and storage of the Urban Heat MiniCubes dataset were made possible with compute resources from the National Science Foundation National Center for Atmospheric Research (NCAR) Computational and Information Systems Laboratory (CISL) and from High Performance Computing at the University of Maryland.

\section*{Author Contributions}
J.S. and M.J.M. conceptualized the dataset. J.S. produced the dataset with M.J.M. guidance. A.K. and A.Y. helped with data download and processing. T.R.H.H. and C.R.H. provided the microwave LST product and provided input on variables to include in the dataset. J.S. wrote the original draft with input from M.J.M. All co-authors contributed to the review and editing of the manuscript.

\section*{Funding}
This material is based upon work supported by the NASA Early Career Research Program (ECIP) award No. 80NSSC24K1044 and the National Science Foundation Graduate Research Fellowship Program under Grant No. DGE 2236417. Any opinions, findings, and conclusions or recommendations expressed in this material are those of the author(s) and do not necessarily reflect the views of the National Science Foundation.

\section*{Supplementary Figures}
\setcounter{table}{0}%
\renewcommand{\thetable}{S\arabic{table}}
\setcounter{figure}{0}%
\renewcommand{\thefigure}{S\arabic{figure}}

\begin{longtable}{|>{\raggedright\arraybackslash}b{5 cm}|b{3 cm}|b{1 cm}|c|}
\caption{List of cities included in the dataset and their Köppen–Geiger climate classification \protect\cite{beck2018present}, projected UTM zone, and GOES satellite used.}\label{city_zones_table}\\
\hline
\textbf{City, Country (ICAO Code)} & \textbf{Climate\newline Classification} & \textbf{UTM Zone} & \textbf{GOES Satellite} \\
\hline
\endfirsthead
Atlanta, Georgia, USA (KATL) & humid subtropical & 16 N & GOES-16\\
\hline
Billings, Montana, USA (KBIL) & semi-arid & 12 N & GOES-16\\
\hline
Bogotá, Colombia (SKBO) & subtropical highland & 18 N & GOES-16\\
\hline
Brasilia, Brazil (SBBR) & tropical savanna & 22 S & GOES-16\\
\hline
Buenos Aires, Argentina (SAEZ) & humid subtropical & 21 S & GOES-16\\
\hline
Cancún, Mexico (MMUN) & tropical wet and\newline dry & 16 N & GOES-16\\
\hline
Caracas, Venezuela (SVMI) & tropical savanna & 19 N & GOES-16\\
\hline
Charlotte, North Carolina, USA (KCLT) & humid subtropical & 17 N & GOES-16\\
\hline
Chicago, Illinois, USA (KORD) & humid continental & 16 N & GOES-16\\
\hline
Dallas, Texas, USA, and Fort Worth, Texas, USA (KDFW) & humid subtropical & 14 N & GOES-16\\
\hline
Denver, Colorado, USA (KDEN) & cool semi-arid & 13 N & GOES-16 \\
\hline
Guadalajara, Mexico (MMGL) & humid subtropical & 13 N & GOES-16\\
\hline
Guatemala City, Guatemala (MGGT) & tropical savanna & 15 N & GOES-16\\
\hline
Havana, Cuba (MUHA) & tropical savanna & 17 N & GOES-16\\
\hline
Houston, Texas, USA (KIAH) & humid subtropical & 15 N & GOES-16\\
\hline
Jacksonville, Florida, USA (KJAX) & humid subtropical & 17 N & GOES-16\\
\hline
La Paz, Bolivia (SLLP) & cold subtropical\newline highland & 19 S & GOES-16\\
\hline
Las Vegas, Nevada, USA (KLAS) & subtropical hot\newline desert & 11 N & GOES-17/18\\
\hline
Lima, Peru (SPJC) & desert & 18 S & GOES-16\\
\hline
Los Angeles, California, USA (KLAX) & semi-arid & 11 N & GOES-17/18\\
\hline
Managua, Nicaragua (MNMG) & tropical wet and\newline dry & 16 N & GOES-16\\
\hline
Manaus, Brazil (SBEG) & tropical monsoon & 20 S & GOES-16\\
\hline
Mexico City, Mexico (MMMX) & subtropical highland & 14 N & GOES-16\\
\hline
Miami, Florida, USA (KMIA) & tropical monsoon & 17 N & GOES-16\\
\hline
Minneapolis, Minnesota, USA (KMSP) & humid continental & 15 N & GOES-16\\
\hline
Monterrey, Mexico (MMMY) & semi-arid & 14 N & GOES-16\\
\hline
Montevideo, Uruguay (SUMU) & humid subtropical & 21 S & GOES-16\\
\hline
Montreal, Quebec, Canada (CYUL) & humid continental & 18 N & GOES-16\\
\hline
New Orleans, Louisiana, USA (KMSY) & humid subtropical & 15 N & GOES-16\\
\hline
New York City, New York, USA (KJFK) & humid subtropical & 18 N & GOES-16\\
\hline
Panama City, Panama (MPTO) & tropical savanna & 17 N & GOES-16\\
\hline
Philadelphia, Pennsylvania, USA (KPHL) & humid subtropical & 18 N & GOES-16\\
\hline
Phoenix, Arizona, USA (KPHX) & hot desert & 12 N & GOES-17/18\\
\hline
Punta Arenas, Chile (SCCI) & subpolar oceanic & 19 S & GOES-16\\
\hline
Quito, Ecuador (SEQM) & subtropical highland & 17 N & GOES-16\\
\hline
Salt Lake City, Utah, USA (KSLC) & humid continental & 12 N & GOES-17/18\\
\hline
San Diego, California, USA, and Tijuana, Mexico (KSAN) & hot-summer Mediterranean & 11 N & GOES-17/18\\
\hline
San Francisco, California, USA, and San Jose, California, USA (KSFO) & warm-summer Mediterranean & 10 N & GOES-17/18\\
\hline
San José, Costa Rica (MROC) & tropical wet and\newline dry & 16 N & GOES-16\\
\hline
San Juan, Puerto Rico (TJSJ) & tropical monsoon & 19 N & GOES-16\\
\hline
Santiago, Chile (SCEL) & cool semi-arid & 19 S & GOES-16\\
\hline
Santo Domingo, Dominican Republic (MDSD) & tropical monsoon & 19 N & GOES-16\\
\hline
São Paulo, Brazil (SBGR) & humid subtropical & 23 S & GOES-16\\
\hline
Seattle, Washington, USA (KSEA) & warm-summer Mediterranean & 10 N & GOES-17/18\\
\hline
St. Louis, Missouri, USA (KSTL) & humid subtropical & 15 N & GOES-16\\
\hline
Tegucigalpa, Honduras (MHTG) & tropical savanna & 16 N & GOES-16\\
\hline
Toronto, Ontario, Canada (CYYZ) & humid continental & 17 N & GOES-16\\
\hline
Washington, District of Columbia, USA, and Baltimore, Maryland, USA (KBWI) & humid subtropical & 18 N & GOES-16\\
\hline
\end{longtable}

\begin{figure}
    \centering
    \includegraphics[width=0.95\linewidth]{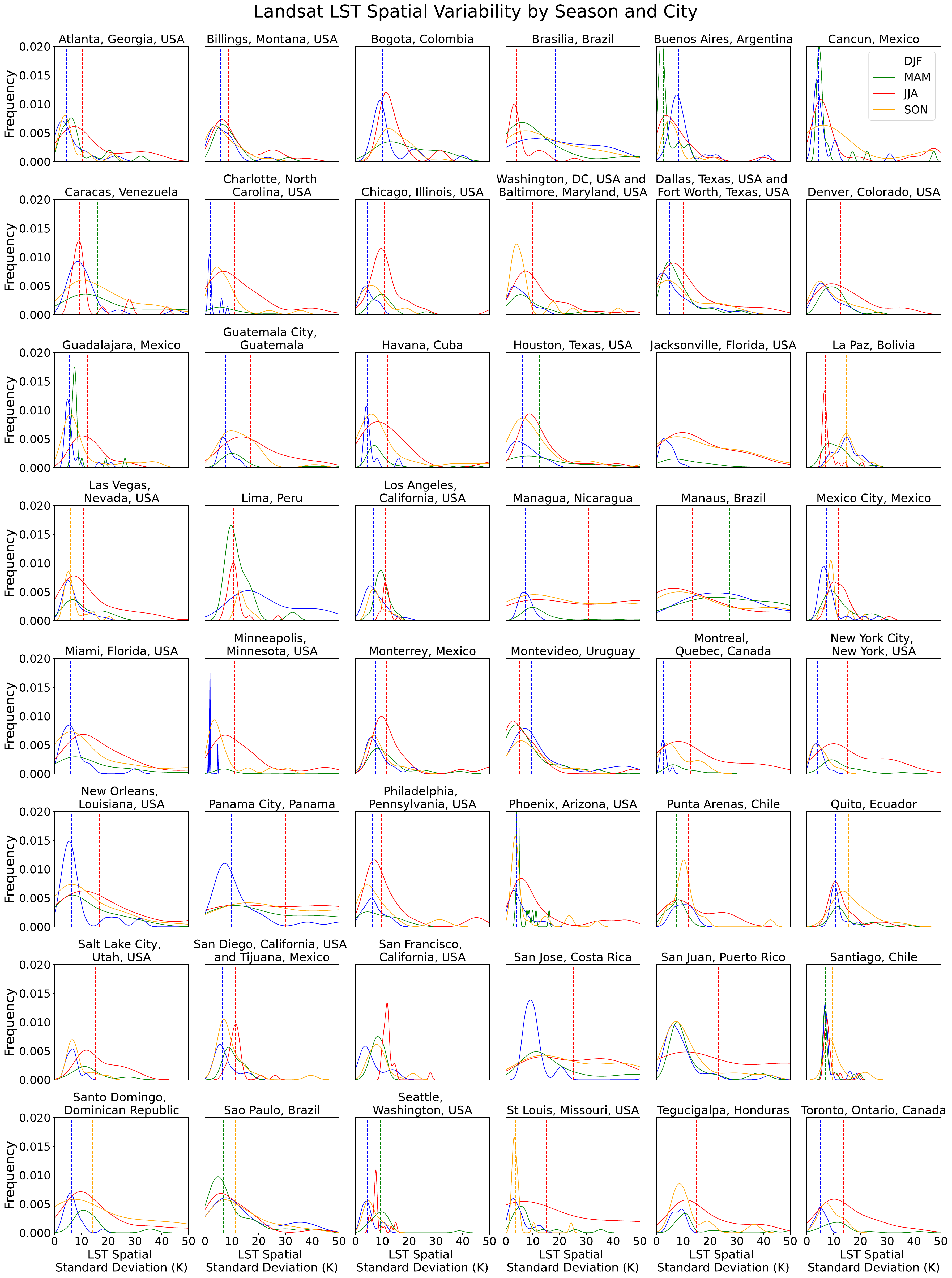}
    \caption{PDFs showing the seasonal variability of the spatial standard deviation of Landsat LST for each city in the dataset. Midpoints of each distribution (CDF $=$ 0.5) are calculated, and dashed vertical lines are plotted at the minimum and maximum values of the corresponding season color to more easily visualize the most and least variable seasons. Symbols next to city names identify exceptions to the expected seasonal pattern: an asterisk marks cities where the highest spatial variability occurs in a season other than summer, while a plus sign marks cities where the lowest spatial variability occurs in a season other than winter.}
    \label{LST std seasonal variability}
\end{figure}

\begin{figure}
    \centering
    \includegraphics[width=0.95\linewidth]{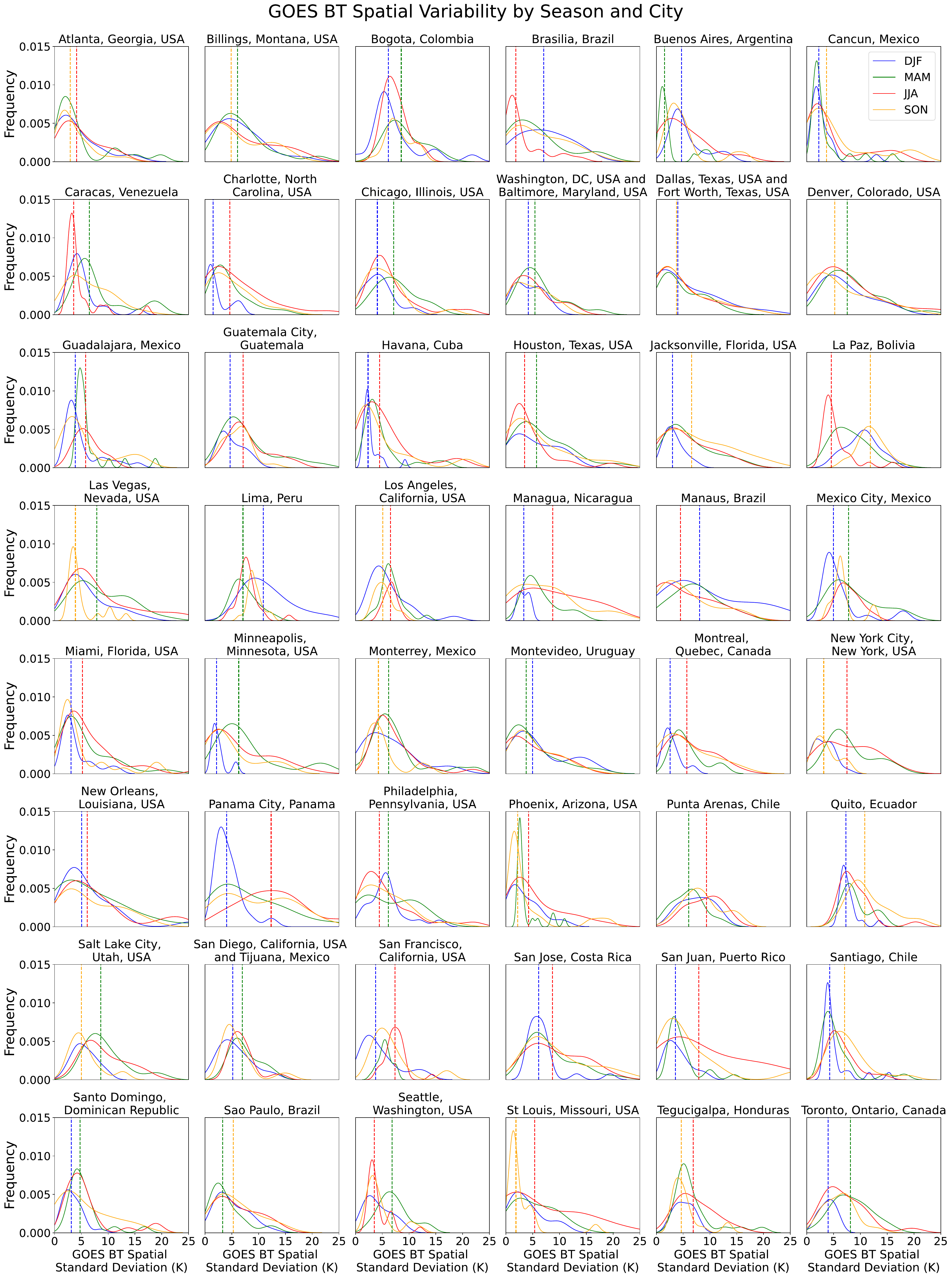}
    \caption{PDFs showing the seasonal variability of the spatial standard deviation of GOES Band 14 BT for each city in the dataset. Data are taken only from the file nearest in time to each Landsat image. Midpoints of each distribution (CDF $=$ 0.5) are calculated, and dashed vertical lines are plotted at the minimum and maximum values of the corresponding season color to more easily visualize the most and least variable seasons. Symbols next to city names identify exceptions to the expected seasonal pattern: an asterisk marks cities where the highest spatial variability occurs in a season other than summer, while a plus sign marks cities where the lowest spatial variability occurs in a season other than winter.}
    \label{GOES BT std seasonal variability}
\end{figure}

\begin{figure}
    \centering
    \includegraphics[width=0.95\linewidth]{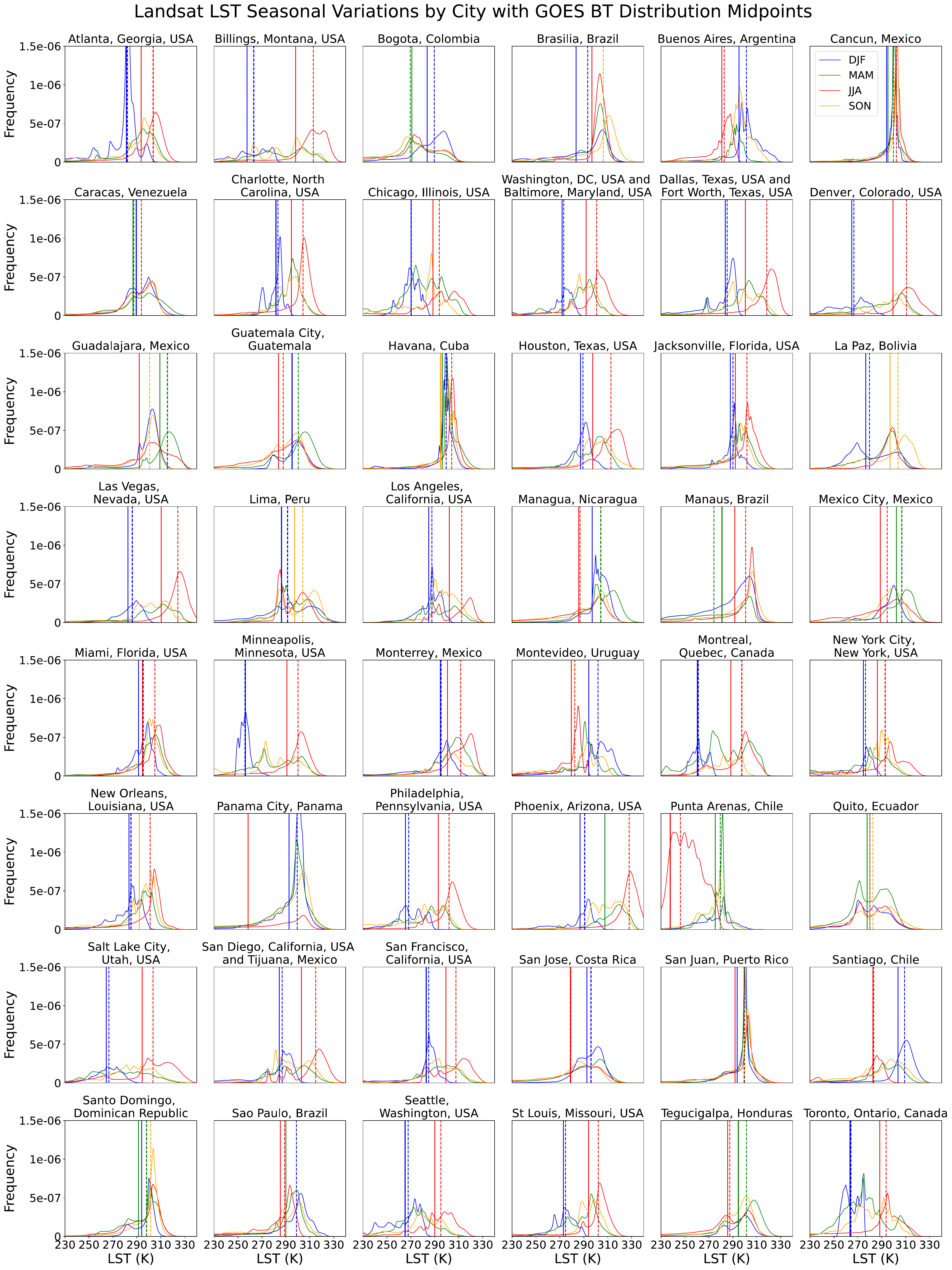}
    \caption{PDFs showing the seasonal variability of Landsat LST for each city in the dataset. Images are coarsened to $750\times750$ pixels via bicubic resampling to speed up computation and are expected to have little effect on the distributional shape. Midpoints of each distribution (CDF $=$ 0.5) are calculated, and dashed vertical lines are plotted at the minimum and maximum values of the corresponding season color to more easily visualize the warmest and coldest seasons. PDFs are calculated, but not plotted, for GOES Band 14 BT, and the minimum and maximum midpoints are plotted as solid lines to facilitate comparisons between Landsat LST and GOES BT values. Asterisks, plus signs, and exclamation points next to city names indicate that Landsat LST and GOES BT coolest seasons do not match, that Landsat LST and GOES BT warmest seasons do not match, and that the GOES BT coolest season temperature is warmer than the Landsat LST coolest season temperature, respectively.}
    \label{LST/BT seasonal variability}
\end{figure}

\begin{figure}
    \centering
    \includegraphics[width=0.95\linewidth]{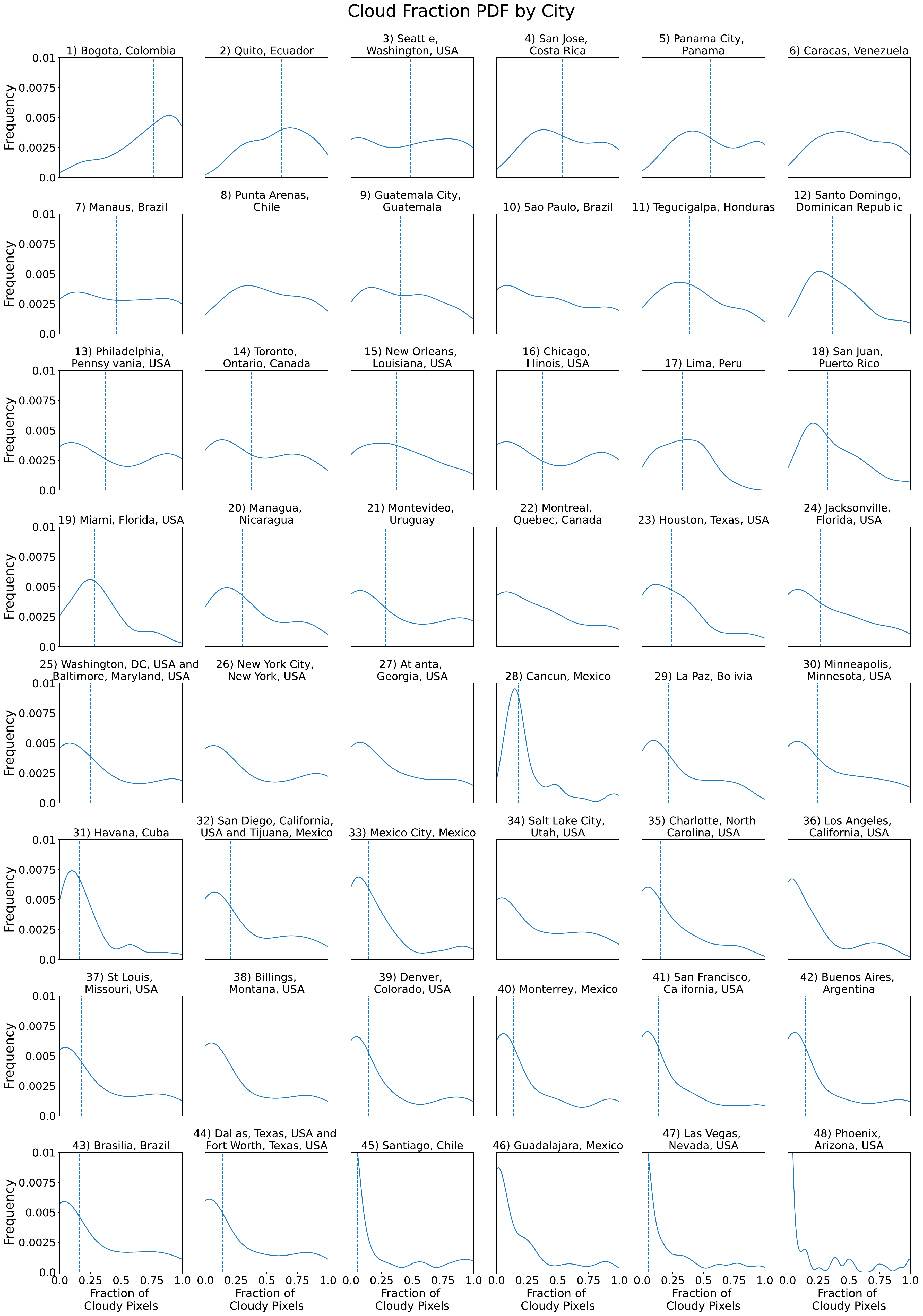}
    \caption{PDFs showing the cloud cover fractions from the Landsat cloud mask for each city in the dataset. Fractions are calculated by dividing the number of pixels in an image where bit 3 (the Cloud bit) equals 1 by the total number of pixels in the image. Midpoints of each distribution (CDF $=$ 0.5) are calculated and plotted as dashed vertical lines. Cities are ordered by the average cloud cover from all of their Landsat files, from cloudiest (1) to least cloudy (48).}
    \label{Cloud cover pdfs}
\end{figure}

\begin{table}[h]
\caption{Architecture used for the convolutional autoencoder neural network used to analyze Landsat LST. BS stands for batch size.} \label{CNN architecture table}
\begin{tabular}{@{}|l|l|l|l|l|l|@{}}
\hline
\textbf{Layer} & \textbf{Features/Channels} & \textbf{Kernel Size} & \textbf{stride} & \textbf{Padding} &  \textbf{Output Size} \\
\hline
Input & 1 & NA & NA & NA & (BS, 1, 32, 32)\\
\hline
Conv2d & 16 & 3 & 1 & 1 & (BS, 16, 32, 32)\\
\hline
BatchNorm2d & 16 & NA & NA & NA & (BS, 16, 32, 32)\\
\hline
ReLU & NA & NA & NA & NA & (BS, 16, 32, 32)\\
\hline
MaxPool2d & NA & 2 & 2 & 0 & (BS, 16, 16, 16)\\
\hline
Conv2d & 32 & 3 & 1 & 1 & (BS, 32, 16, 16)\\
\hline
BatchNorm2d & 32 & NA & NA & NA & (BS, 32, 16, 16)\\
\hline
ReLU & NA & NA & NA & NA & (BS, 32, 16, 16)\\
\hline
MaxPool2d & NA & 2 & 2 & 0 & (BS, 32, 8, 8)\\
\hline
Conv2d & 64 & 3 & 1 & 1 & (BS, 64, 8, 8)\\
\hline
BatchNorm2d & 64 & NA & NA & NA & (BS, 64, 8, 8)\\
\hline
ReLU & NA & NA & NA & NA & (BS, 64, 8, 8)\\
\hline
MaxPool2d & NA & 2 & 2 & 0 & (BS, 64, 4, 4)\\
\hline
Conv2d & 64 & 3 & 1 & 1 & (BS, 64, 4, 4)\\
\hline
BatchNorm2d & 64 & NA & NA & NA & (BS, 64, 4, 4)\\
\hline
ReLU & NA & NA & NA & NA & (BS, 64, 4, 4)\\
\hline
MaxUnpool2d & NA & 2 & 2 & 0 & (BS, 64, 8, 8)\\
\hline
Conv2d & 32 & 3 & 1 & 1 & (BS, 32, 8, 8)\\
\hline
BatchNorm2d & 32 & NA & NA & NA & (BS, 32, 8, 8)\\
\hline
ReLU & NA & NA & NA & NA & (BS, 32, 8, 8)\\
\hline
MaxUnpool2d & NA & 2 & 2 & 0 & (BS, 32, 16, 16)\\
\hline
Conv2d & 16 & 3 & 1 & 1 & (BS, 16, 16, 16)\\
\hline
BatchNorm2d & 16 & NA & NA & NA & (BS, 16, 16, 16)\\
\hline
ReLU & NA & NA & NA & NA & (BS, 16, 16, 16)\\
\hline
MaxUnpool2d & NA & 2 & 2 & 0 & (BS, 16, 32, 32)\\
\hline
Conv2d & 1 & 3 & 1 & 1 & (BS, 1, 32, 32)\\
\hline
\end{tabular}
\end{table}

\end{document}